\documentclass[letterpaper,titlepage,11pt]{article}
\usepackage{hyperref}
\usepackage{amssymb,amsmath,amsfonts}
\usepackage{epsfig}
\usepackage{graphicx}
\usepackage{float}
\usepackage[english]{babel}
\usepackage[T1]{fontenc}
\usepackage[applemac]{inputenc}
\def\clock{{\count0=\time
           \divide\count0 60
           \ifnum\count0<10 0\fi\the\count0
           \multiply\count0 -60 \advance\count0 \time
           :\ifnum\count0<10 0\fi \the\count0
         }}
\newcommand{\timestamp}{{\small\vbox{\hbox{\tt\jobname.tex}
\hbox{\the\day/\the\month/\the\year, \clock}}}}

\newcommand{\CO}{\mathcal{O}}
\newcommand{\CN}{\mathcal{N}}
\newcommand{\CI}{\mathcal{I}}

\newcommand{\CM}{\mathcal{M}}

\newcommand{\Z}{\mathbb{Z}}
\newcommand{\C}{\mathbb{C}}
\newcommand{\R}{\mathbb{R}}

\newcommand{\spa}{\ , \ \ }

\newcommand{\ds}{\displaystyle}

\newtheorem{definition}{Definition}[section]

\newtheorem{theorem}[definition]{Theorem}

\newcommand{\squ}{\noindent $\square$}

\numberwithin{equation}{section}

\begin{document}

\begin{titlepage}

\begin{flushright}
NORDITA-2011-95\\
\end{flushright}

\vskip 1.7cm

\centerline{\huge \bf Domain Structure of Black Hole Space-Times}
\vskip 0.3cm
\centerline{\huge \bf with a Cosmological Constant}
\vskip 1.7cm
\centerline{\bf  Jay Armas$^1$, Pawe\l \space Caputa$^{1,2}$ and Troels Harmark$^3$}
\vskip 0.8cm
\centerline{\sl $^1$ The Niels Bohr Institute, Copenhagen University}
\centerline{\sl Blegdamsvej 17, DK-2100 Copenhagen \O, Denmark}
\vskip 0.3cm
\centerline{\sl $^2$ The Niels Bohr International Academy}
\centerline{\sl Blegdamsvej 17, DK-2100 Copenhagen \O, Denmark}
\vskip 0.3cm
\centerline{\sl $^3$ NORDITA}
\centerline{\sl Roslagstullsbacken 23,
SE-106 91 Stockholm,
Sweden}

\vskip 1cm

\centerline{\small\tt jay@nbi.dk, caputa@nbi.dk, harmark@nordita.org}

\vskip 1.5cm

\centerline{\bf Abstract} \vskip 0.2cm \noindent We generalize the domain structure for stationary black hole space-times to include asymptotically de Sitter and Anti-de Sitter space-times. Given a set of commuting Killing vector fields of a space-time the domain structure lives on the submanifold of the orbit space on which at least one of the Killing vector fields has zero norm. In general the domain structure provides topological and geometrical invariants of  black hole space-times that in specific cases have proven to be a crucial part of a full characterization leading to uniqueness theorems. In four and five dimensions the domain structure generalizes the rod structure. We examine in detail the domain structure for four, five, six and seven-dimensional black hole space-times including a very general class of spherically symmetric and static black hole space-times as well as the exact solutions for Kerr-(Anti)-de Sitter black holes. While for asymptotically Anti-de Sitter space-times the domain structures resemble that of asymptotically flat space-times, the domain structures of asymptotically de Sitter space-times are shown to be compact. Finally, we find under certain assumptions that the horizon topologies for four- and five-dimensional black holes are restricted by our domain structure analysis.

\end{titlepage}

\small
\tableofcontents
\normalsize
\setcounter{page}{1}


\section{Introduction}

The celebrated uniqueness theorems for four-dimensional asymptotically flat and stationary black holes state that there is at most one possible black hole space-time available given the asymptotically measured conserved charges, namely the mass, angular momentum and electric and magnetic charges \cite{Israel:1967wq,Hawking:1972vc}.  In this sense black holes are in stark contrast to distributions of matter: For matter distributions one would need an infinite number of multipole moments to find the metric that the matter distribution sources. The uniqueness of black holes thus seems a very special property which one would assume says something deep about their nature. However, one could also imagine that it is a coincidence, $i.e.$ an artifact of some special property of this particular family of space-times which is lost if one considers a larger family of black hole space-times. The question is thus whether uniqueness is an essential part of black hole physics or a just a contingent feature?

One way to consider a larger family of black hole space-times is to include asymptotically de Sitter (dS) and Anti-de Sitter (AdS) black holes, $i.e.$ black hole space-times for which the Einstein equations include a cosmological constant. Contrary to the asymptotically flat case, the results for asymptotically dS and AdS black holes are still very few. In the case of static black hole space-times it has been shown that the Schwarzschild-dS and Schwarzschild-AdS black holes are perturbatively unique \cite{Ishibashi:2003ap}. Moreover, a static and asymptotically AdS black hole space-time solving the Einstein equations with a cosmological constant is uniquely given by the Schwarzschild-AdS black hole solution \cite{Anderson:2002xb}.%
\footnote{This includes a technical assumption of a $\C^2$ conformal spatial completion of the space-time.}
 However, there are no uniqueness theorems for stationary asymptotically dS or AdS black holes. One reason for the lack of progress on this front is that the uniqueness theorems for stationary asymptotically flat black holes rely on certain special properties of the Einstein equations that are absent when introducing a cosmological constant. 

Another way to consider a larger family of black hole space-times is to include asymptotically flat black hole space-times with more than four space-time dimensions.%
\footnote{Another way is to consider black hole space-times which are asymptotically Kaluza-Klein space-times and generalizations thereof, see for example the review \cite{Harmark:2007md}.}
In this case it has become clear, in particular for the black hole space-times which solve the vacuum Einstein equations, that there are many possible black hole space-times given the asymptotically measured conserved charges \cite{Myers:1986un,Emparan:2001wn,Emparan:2007wm,Dias:2009iu}. In five dimensions the discovery of the black ring \cite{Emparan:2001wn,Pomeransky:2006bd}, in addition to the Myers-Perry black hole \cite{Myers:1986un}, shows explicitly that it is not enough to specify the mass and angular momenta in order to uniquely characterize the black hole space-time. Moreover, it is not even enough if we in addition specify the horizon topology. However, it was shown in \cite{Harmark:2004rm} that all five-dimensional black holes solving the vacuum Einstein equations possess the so-called rod-structure provided the black hole has two rotational Killing vector fields.%
\footnote{The definition of the rod-structure in \cite{Harmark:2004rm} generalizes the rod structure defined for black hole space-times with $D-2$ orthogonal commuting Killing vector fields in \cite{Emparan:2001wk}. See \cite{Kleihaus:2009dm,Kleihaus:2010pr} for a different type of generalization of the rod-structure.} In addition to topological invariants, the rod-structure also includes geometrical invariants of the black hole space-time. Including these extra invariants of the rod-structure it was shown in \cite{Hollands:2007aj,Hollands:2008fm} that a stationary, asymptotically flat, five-dimensional black hole space-time solving the vacuum Einstein equations, and with two rotational Killing vector fields, is unique when specifying the mass, angular momenta and the rod-structure.%
\footnote{This was generalized to include disconnected horizons in \cite{Armas:2009dd} with the extra ingredient that one should specify the angular momenta for each connected component of the event horizon. Further generalizations to five-dimensional black holes in five-dimensional Einstein-Maxwell and Einstein-Maxwell-Chern-Simons gravity can be found in \cite{Hollands:2007qf,Tomizawa:2009ua,Tomizawa:2009tb,Armas:2009dd}. See \cite{Ida:2011jm} for a review.}

As a consequence of including the five-dimensional black holes in the family of black hole space-times we are forced to revise the concept of uniqueness of black holes that we know from four dimensions. Instead of only considering the asymptotically measured conserved charges, we should broaden our definition of uniqueness of black holes and think more generally in terms of invariants of the black hole space-time. There are three types of invariants: 1)~Topological invariants of the black hole space-time, 2)~Geometrical invariants of the black hole space-time, and
3)~Physical quantities measured either locally for each event horizon, or globally, meaning in the asymptotic region ($e.g.$ mass, angular momenta, charges, etc.).
For the class of five-dimensional black holes discussed above the topological and geometrical invariants are given by the rod-structure, with the topological invariants defined by how the rods are split in intervals and which directions the rods have in the vector space spanned by the rotational Killing vector fields, while the geometrical invariants are the lengths of the rods \cite{Harmark:2004rm,Hollands:2007aj,Hollands:2008fm}.

The natural question is then whether the rod-structure can be generalized to a larger family of black hole space-times. Indeed, the uniqueness theorems for asymptotically flat space-times in four and five dimensions all rely on very specific properties of Einstein equations that do not hold in higher dimensions nor for space-times with a cosmological constant. 

However, in \cite{Harmark:2009dh} a natural generalization of the rod structure invariants is proposed. The new set of invariants, known as the {\sl domain structure}, is defined given a stationary black hole space-time with any number of space-time dimensions and any number of commuting Killing vector fields. The domain structure lives on the submanifold of the orbit space where at least one of the Killing vector fields has zero norm. Depending on which Killing vector field has zero norm the submanifold is naturally divided into domains. A domain corresponds either to a set of fixed points of a spatial symmetry, such as a rotational symmetry of the space-time, or to a Killing horizon, depending on whether the characterizing Killing vector field is space-like or time-like near the domain.%
\footnote{In \cite{Hollands:2010qy} the invariants for the special case of five-dimensional asymptotically flat black hole space-times with a single rotational Killing vector field are considered in detail. This includes the domain structure invariants.} 

In \cite{Harmark:2009dh} the domain structure was developed for asymptotically flat space-times and explored in detail in the six and seven-dimensional cases with the maximal possible number of rotational Killing vector fields. It is important to emphasize that unlike in the original definition of the rod-structure in \cite{Harmark:2004rm,Hollands:2007aj,Hollands:2008fm} the domain structure is defined without using any knowledge of how the black hole space-time solves the Einstein equations. For asymptotically flat black holes in four and five dimensions this means that the domain structure straightforwardly generalizes the rod structure from solutions of the vacuum Einstein equations to theories with any type of matter present. In addition this makes it possible to define the domain structure on non-asymptotically flat spaces such as space-times with a cosmological constant. This is what we explore in this paper.

In this paper we define the domain structure for asymptotically dS and AdS black hole space-times. This way we show that the set of invariants for black hole space-times, originally introduced with the rod structure in \cite{Harmark:2004rm}, can be naturally extended to a set of invariants for black hole space-times with a cosmological constant for any number of space-time dimensions and any number of commuting Killing vector fields. Since with this work we are able to define the domain structure invariants for a very large family of black hole space-times, we believe this can help in paving out a possible way of how to generalize what one means with uniqueness of black holes. 

We begin this paper with reviewing and generalizing the domain structure invariants of \cite{Harmark:2009dh} in Section \ref{sec:generaldom} for the particular case of $D$-dimensional stationary black hole space-times with $D-2$ commuting Killing vector fields. In this case the domain structure becomes one-dimensional with the domains being intervals and with each interval associated with a linear combination of the commuting Killing vector fields. We notice in particular that the geometrical invariants, which in this case are the lengths of the intervals, can be invariantly defined as long as the chosen Killing vector field associated with time-translation is properly defined with respect to a reference space-time in the asymptotic region. 

In Section \ref{sec:asymptads} we investigate the domain structure of four- and five-dimensional stationary asymptotically AdS black hole space-times, in five dimensions with two rotational Killing vector fields. We find the domain structure for pure AdS$_4$ and AdS$_5$ and go on to consider the domain structure for a very general class of asymptotically AdS spherically symmetric black hole space-times. Furthermore, we consider the domain structure of Kerr-Newman-AdS$_4$ and Kerr-AdS$_5$ space-times, and we consider furthermore the domain structure of black rings in AdS$_5$. We notice that the domain structure of asymptotically AdS$_4$ and AdS$_5$ black hole space-times lives on an infinitely long line and hence closely resembles that of asymptotically flat black hole space-times.

In Section \ref{sec:asymptds} we investigate the domain structure of four- and five-dimensional stationary asymptotically dS black hole space-times with the maximal number of commuting rotational Killing vector fields. We find the domain structure for pure dS$_4$ and dS$_5$ and subsequently for a very general class of asymptotically dS spherically symmetric black hole space-times. Furthermore, we consider the domain structure of Kerr-Newman-dS$_4$ and Kerr-dS$_5$ space-times, and we briefly consider the domain structure for a pair of black holes in dS$_4$ as well as  for black rings in dS$_5$. In general we find that the domain structure of asymptotically dS$_4$ and dS$_5$ black hole space-times lives on a compact circle, unlike the cases of asymptotically flat or asymptotically AdS black hole space-times.

In the above-mentioned sections we also find that assuming that the domain structure space is connected we can infer from the domain structure of stationary black hole space-times with the maximal number of commuting rotational Killing vector fields that the possible horizon topologies are restricted to be $S^2$ for four-dimensional black holes and $S^3$, $S^2 \times S^1$ or Lens-space for five-dimensional black holes. This holds for asymptotically flat, dS and AdS space-times, thus generalizing the analysis of \cite{Hollands:2007aj} in the five-dimensional asymptotically flat case.

In Section \ref{sec:higherdim} we generalize the domain structure to the case of solutions with less than $D-2$ Killing vector fields, considering in detail the case of $D-3$ Killing vector fields. We apply this analysis to six and seven dimensional stationary and asymptotically dS black hole space-times with the maximal possible number of commuting rotational Killing vector fields. We find that the domain structure lives on a two-dimensional sphere.

In Section \ref{sec:concl} we end with a discussion and outlook on future directions of research.

\section{Domain structure of black hole space-times with $D-2$ Killing vector fields}
\label{sec:generaldom}

In this section we explain the general idea of the domain structure \cite{Harmark:2009dh} of stationary $D$-dimensional black hole space-times in the special case of $p=D-2$ commuting Killing vector fields and review in particular how it works for asymptotically flat space-times.  In Section \ref{sec:higherdim} we consider briefly space-times with $p$ commuting Killing vector fields where $p < D-2$.

Consider a $D$-dimensional space-time with $p=D-2$ commuting linearly independent Killing vector fields. In detail we are given a $D$-dimensional manifold $\CM_D$ with a Lorentzian signature metric with $p=D-2$ commuting linearly independent Killing vector fields $V_{(i)}$, $i=0,1,...,p-1$. The Killing vector fields are such that they generate the isometry group $\R \times U(1)^{p-1}$. In particular the $p-1$ $U(1)$ symmetries are generated by the $p-1$ space-like Killing vector fields $V_{(i)}$, $i=1,...,p-1$, while the Killing vector field $V_{(0)}$ generates the $\R$ isometry. 

We normalize the periods of the $U(1)$ flows of $V_{(i)}$, $i=1,...,p-1$ to be $2\pi$. There are restrictions on choosing another basis for the Killing vector fields generating the $U(1)^{p-1}$ group. A new basis $V_{(i)}'$, $i=1,...,p-1$ is in
general a linear combination
\begin{equation}
V_{(i)}' = \sum_{j=1}^{p-1} U_{ij} V_{(j)}
\end{equation}
Since we demand each $V_{(i)}'$ Killing vector field to generate a $U(1)$ flow with period $2\pi$ this restricts the above transformation such that $U \in GL(p-1,\Z)$ and $\det U = \pm 1$. In particular, the rows of $U$ are mutually prime numbers.

As shown in \cite{Harmark:2009dh} we can locally find coordinates $x^0,x^1,...,x^{p-1},r,z$ such that the Killing vectors take the form
\begin{equation}
\label{killv} V_{(i)} = \frac{\partial}{\partial x^i}
\end{equation}
for $i=0,1,...,p-1$ and the metric takes the form
\begin{equation}
\label{genmetric}
\begin{array}{c} \ds ds^2 = G_{ij} (dx^i + A^i) (dx^j + A^j) +
e^{2\nu} ( dr^2 + \lambda^2 dz^2 )   \spa r = \sqrt{ | \det G_{ij} | } 
\end{array}
\end{equation}
which we in this paper dub the {\sl canonical form of the metric}.%
\footnote{In \cite{Harmark:2009dh} the canonical form of the metric included a condition on the behavior of $\lambda$ for $r \rightarrow \infty$. We explain below why this condition is not necessary for defining the domain structure.} We have introduced the one-forms $A^i = A^i_r dr + A^i_z dz$, and all the components of the metric only depend on $(r,z)$, $i.e.$ the metric is written in terms of the functions $G_{ij}(r,z)$, $A^i_r(r,z)$, $A^i_z (r,z)$, $\nu (r,z)$ and $\lambda(r,z)$. Note that we demand without loss of generality that $\lambda \geq 0$.

Consider all coordinate transformations that preserve the canonical form \eqref{genmetric} of the metric. First there are rigid transformations of the $(x^0,x^1,...,x^{p-1})$ coordinates. These are constrained by the above condition on the $U(1)$ flows of the $p-1$ Killing vector fields $V_{(i)}$, $i=1,...,p-1$. Secondly, there are transformations of the type $x^i \rightarrow x^i - \alpha^i(r,z)$. These act as gauge transformations of the one-forms $A^i$ and do not affect other components of the metric. Finally, there are transformations of the $(r,z)$ coordinates. Consider a transformation $\tilde{r} = \tilde{r} (r,z)$ and $\tilde{z} = \tilde{z} (r,z)$. We first observe that $\tilde{r} (r,z) =r$ since $r^2 = |\det G_{ij} |$. Using $g^{rz} = 0$ we find $g^{\tilde{r} \tilde{z}} = e^{-2\nu} ( \partial \tilde{z} / \partial r )$ hence $\tilde{z}$ cannot depend on $r$. Thus the most general coordinate transformation of the $(r,z)$ coordinates is $\tilde{r} = r$ and $\tilde{z} = \tilde{z} (z)$. Note here that we restrict ourselves to coordinate transformations which are one-to-one in the given patch of the space-time that one is considering. Locally we impose this by demanding $d\tilde{z}/dz > 0$.

From the $D$-dimensional space-time $\CM_D$ we define the two-dimensional manifold $\CN$ as the quotient space $\CM_D / \sim$ where the equivalence relation $\sim$ is such that two points in $\CM_D$ are equivalent if they can be connected by an integral curve of a linear combination of the Killing vector fields $V_{(i)}$, $i=0,1,....,p-1$. This is known as the orbit space. Using the metric \eqref{genmetric} we parameterize $\CN$ with the coordinates $(r,z)$ and equip it with the metric
\begin{equation}
\label{Nmetric}
ds_{\CN}^2 = dr^2 + \lambda^2 dz^2
\end{equation}
The domain structure of the space-time $\CM_D$ is defined on the one-dimensional submanifold of $\CN$
\begin{equation}
B = \{ q \in \CN | \det G = 0 \}
\end{equation}
We see that in the $(r,z)$ parameterization of $\CN$ this submanifold is defined by $r=0$. Employing the metric \eqref{Nmetric} we can therefore introduce the metric
\begin{equation}
\label{Bmetric}
ds_{B}^2 = \lambda^2 |_{r=0} \, dz^2
\end{equation}
on the one-dimensional set $B$. Note that in general we cannot assume that $B$ is a connected set. Define the sets
\begin{equation}
Q_k = \{ q \in \CN | \dim \ker G \geq k \}
\end{equation}
We see that $Q_0 = \CN$ and $Q_1 = B$. Considering the set $Q_2$ it is shown in \cite{Harmark:2009dh} that this is a set of isolated points when assuming the absence of curvature singularities and in particular conical singularities in the space-time. Under the same assumption it is furthermore shown that the vectors in $\ker G$ are constant in the connected pieces of $B-Q_2$. Using this we define for a point $q \in B - Q_2$ the {\sl domain} $D$ containing $q$ as the maximal connected subset of $B$ such that $q \in D$ and such that for any point $q' \in D$ we have $\ker G (q)\subset \ker G (q')$. Using this we can consider all the domains $D_1, D_2, ... , D_N$ for the given space-time $\CM_D$ (we assume here that the number of domains is finite). Then we have $D_i \cap D_j \subset Q_2$ for $i \neq j$ and $B = D_1 \cup D_2 \cup \cdots \cup D_N$. We now have the theorem \cite{Harmark:2009dh}

\begin{theorem}
\label{theo:dom} Let $D_1,...,D_N$ be the domains of $B$. We have
that $B = D_1 \cup D_2 \cup \cdots \cup D_N$. For each domain $D_m$
we can find a Killing vector field $W_m$ such that $W_m \in \ker G$
for all points in $D_m$. We call $W_m$ the {\sl direction} of the
domain $D_m$. If $W_m$ is space-like for $r\rightarrow 0$ we can
write it in the form
\begin{equation}
\label{Wmspace}
W_m = \sum_{i=1}^{p-1} q_i V_{(i)}
\end{equation}
where the $q_i$'s are relatively prime numbers. Then $W_m$ generates
a $U(1)$ isometry and the generated flow has period $2\pi$. In this
case we say that the direction $W_m$ is space-like.

If $W_m$ is time-like for $r\rightarrow 0$ we can write it in the
form
\begin{equation}
\label{Wmtime}
W_m = V_{(0)} +  \sum_{i=1}^{p-1} \Omega_i V_{(i)}
\end{equation}
and the domain $D_m$ is a Killing horizon for the Killing vector field $W_m$. In this case we say that the direction $W_m$ is time-like.
\squ
\end{theorem}

In the parameterization of the canonical form of the metric \eqref{genmetric} the domains are intervals in the $z$ coordinate. Thus we can write $D_m = [ a_m , a_{m+1} ]$, $m=1,...,N$, where all the points $(r,z) = (0,a_1), ... , (0,a_{N+1})$ are in $Q_2$. Using this with the metric \eqref{Bmetric} we define the length of the domain $D_m$ as
\begin{equation}
\label{domlength}
| D_m | = \int_{a_m}^{a_{m+1}} \lambda (0, z) dz
\end{equation}
A crucial question is whether this length is invariant under coordinate transformations that preserve the canonical form of the metric \eqref{genmetric}. First we note that it is trivially invariant under transformations $x^i \rightarrow x^i - \alpha^i (r,z)$. Consider then transformations of the $(r,z)$ coordinates. Above we have shown that the most general transformation is $\tilde{r} = r$ and $\tilde{z} = \tilde{z} (z)$. Under such a transformation $\lambda$ is transformed to $\tilde{\lambda} = (d\tilde{z} / dz )^{-1} \lambda$ hence we have $\lambda dz = \tilde{\lambda} d \tilde{z}$. Thus, the length of a domain as defined by \eqref{domlength} is invariant under such coordinate transformations. Finally, we consider the rigid transformations of the $(x^0,x^1,...x^{p-1})$ coordinates. Under such transformations $| \det G_{ij} |$ can change and thus the $r$ coordinate changes which affects the overall scale of the $(r,z)$ coordinates and thus also the domain length \eqref{domlength}. Therefore, the domain length \eqref{domlength} is defined with respect to the choice of Killing vector fields $V_{(i)}$, $i=0,1,...,p-1$, of the space-time. However, as explained above, the choice of the $p-1$ Killing vector fields $V_{(i)}$, $i=1,...,p-1$, generating the $U(1)^{p-1}$ isometry, is restricted by setting the periods of the $U(1)$ flows to be $2\pi$. This means that different choices of the $p-1$ Killing vector fields $V_{(i)}$, $i=1,...,p-1$, do not change $| \det G_{ij} |$ and hence preserve the domain length \eqref{domlength}. Thus, we can conclude that the domain length \eqref{domlength} is measured solely with respect to the choice of the Killing vector field $V_{(0)}$ generating the $\R$ isometry.

In order to make the domain length \eqref{domlength} an invariant of black hole space-times we should specify how to choose $V_{(0)}$. This is done by requiring that the space-time has an asymptotic region where the space-time approaches a reference space-time, such as $D$-dimensional Minkowski space, AdS or dS. The $V_{(0)}$ of the given space-time should then approach the $V_{(0)}$ of the reference space-time in the asymptotic region. $V_{(0)}$ in the reference space-time corresponds to the time-translation Killing vector field. That we need to specify how to choose $V_{(0)}$ is physically due to the fact that a length has to be measured with respect to a reference space-time. Thus, like the physical quantities of a black hole such as the mass or the surface gravity, the domain length \eqref{domlength} is a quantity which is measured by an asymptotic observer. The choice of $V_{(0)}$ is thus a standard part of choosing the "measurement rods" for a given black hole space-time and it is necessary for defining any of the well-established physical quantities for black hole space-times. We review the standard choice of $V_{(0)}$ for asymptotically flat space-times below and for asymptotically AdS and dS space-times in Sections \ref{sec:asymptads} and \ref{sec:asymptds}, respectively.

As an addendum to the above we note that while the domain length \eqref{domlength} obviously is sensitive to rescalings of $V_{(0)}$ it is not sensitive to the addition of the space-like Killing vector fields $V_{(i)}$, $i=1,2,...,p-1$. Consider a linear transformation
\begin{equation}
V_{(0)}' = V_{(0)} + \sum_{i=1}^{p-1} u_i V_{(i)}
\end{equation}
with $V_{(i)}' = V_{(i)}$, $i=1,2,...,p-1$. Such a transformation preserves the determinant $\det G_{ij}$ and hence the domain length.

We are now ready to define the {\sl domain structure} for a given $D$-dimensional black hole space-time with $p=D-2$ commuting and linearly independent Killing vector fields $V_{(i)}$, $i=0,1,...,p-1$, under the restrictions given in the beginning of this section. The domain structure of the black hole space-time is defined as the split-up of $B$ in domains $B = D_1 \cup D_2 \cup \cdots \cup D_N$ along with the directions $W_m$, $m=1,2,...,N$, of the domains and the lengths of the domains $|D_1|,|D_2|,...,|D_N|$ as defined by \eqref{domlength}. Note that the split-up of $B$ in domains is defined up to coordinate transformations of $z$ as explained above.
Thus, the domain structure defines a collection of topological invariants of the space-time in the form of the split-up of $B$ in domains $B = D_1 \cup D_2 \cup \cdots \cup D_N$ and with the directions $W_m$ of the space-like domains, along with a collection of geometrical invariants of the space-time in the form of the lengths of the domains $|D_1|,|D_2|,...,|D_N|$. In particular one can read off the topology of an event horizon in the space-time using the topological invariants. 

It is important to note that the domain structure is well-defined irrespective of how the space-time in question solves the Einstein equations. In particular we do not need to know what matter fields are present in the Einstein equations. All the information that we require about the space-time concerns the Killing vector fields.

The form of the metric \eqref{genmetric} is considered in particular for stationary asymptotically flat black hole space-times in \cite{Harmark:2009dh}.%
\footnote{It is also considered for asymptotically Kaluza-Klein space-times in \cite{Harmark:2009dh}.} For asymptotically flat space-times
we take $V_{(0)}$ to be the Killing vector field asymptoting to the time-translation Killing vector field of $D$-dimensional Minkowski space in the asymptotic region. This means in particular that in the asymptotic region, where we are arbitrarily close to spatial infinity, $V_{(0)}$ is orthogonal to all $V_{(i)}$, $i=1,2,...,p-1$, and the norm squared $g_{\mu\nu} V_{(0)}^\mu V_{(0)} ^\nu$ of $V_{(0)}$ approaches $-1$. Employing this we thus have a unique definition of the lengths of the domains from \eqref{domlength}. With respect to writing the metric of the space-time in the canonical form \eqref{genmetric} we can use a single coordinate patch to describe the metric from the asymptotic region and all the way to the event horizon(s). For $r \rightarrow \infty$ we reach (part of) spatial infinity of the space-time. Although it is not necessary for measuring the domain lengths \eqref{domlength} we add in \cite{Harmark:2009dh} the extra requirement on the canonical form of the metric for asymptotically flat black hole space-times that $\lambda \rightarrow 1$ for $r \rightarrow \infty$ since this is a natural generalization of the canonical form of the metric given in \cite{Harmark:2004rm}. One then has that the $z$ coordinate ranges from $-\infty$ to $\infty$. The domains are thus the intervals $D_m = [a_m,a_{m+1}]$ with $a_1 = -\infty$, $a_{N+1} = \infty$ and $a_m < a_{m+1}$.%
\footnote{Note that this assumes $B$ is connected.} Note that for the end domains $D_1 = ( -\infty , a_2 ]$ and $D_N = [ a_N , \infty )$ it follows from the above that the points $a_1 = -\infty$ and $a_{N+1} = \infty$ are not included.

For the special case of asymptotically flat metrics solving the vacuum Einstein equations $R_{\mu\nu}=0$ the domain structure reduces to the rod structure defined in \cite{Harmark:2004rm,Hollands:2007aj,Hollands:2008fm}.%
\footnote{We do not use the name rod structure here since that name entails that the domains should be interpretable as sources as in \cite{Emparan:2001wk} which means that one should have knowledge of how the metric solves the Einstein equations and this is not necessary to define the domain structure.} Furthermore, it reduces to the other manifestations of the rod/domain structure found for asymptotically flat black hole space-times which are solutions of five-dimensional Einstein-Maxwell and Einstein-Maxwell-Chern-Simons \cite{Hollands:2007qf,Tomizawa:2009ua,Tomizawa:2009tb,Armas:2009dd}.

In the rest of this paper we employ the above construction of the domain structure to static and stationary asymptotically AdS and dS black hole space-times. Since we require $p=D-2$ commuting Killing vector fields we can treat either four-dimensional static/stationary asymptotically AdS and dS black holes space-times with one rotational Killing vector field, or five-dimensional static/stationary asymptotically AdS and dS black holes space-times with two rotational Killing vector fields. However, we briefly consider higher-dimensional cases in Section \ref{sec:higherdim}.

\section{Domain structure of asymptotically AdS space-times}
\label{sec:asymptads}

In this section we first consider the Canonical coordinates and the domain structure of AdS$_4$ and AdS$_5$. Subsequently we employ this to find the domain structure of asymptotically AdS$_4$ and AdS$_5$ stationary black hole space-times.

\subsection{Domain structure of AdS}
\label{sec:domads}

In this section we study the domain structure of the AdS$_4$ and AdS$_5$ space-times. 

\subsubsection*{Domain structure of AdS$_4$}

The metric of AdS$_4$ in the global patch can be written as
\begin{equation}
\label{ads4}
ds^2 = - f dt^2 + \frac{d\rho^2}{f} + \rho^2 ( d\theta^2 + \sin^2 \theta d\phi^2 ) \spa f (\rho)= 1 + \frac{\rho^2}{L^2}
\end{equation}
Here $L$ is the length-scale of AdS$_4$ in terms of which the cosmological constant is $\Lambda = - 3/ L^2$.
The ranges of the $(\rho,\theta)$ coordinates are $\rho \geq 0$ and $0\leq \theta \leq \pi$. In terms of the above metric AdS$_4$ has the two commuting Killing vector fields 
\begin{equation}
\label{ads4killing}
V_{(0)} = \frac{\partial}{\partial t} \spa V_{(1)} =  \frac{\partial}{\partial \phi}
\end{equation}
We choose to define the domain structure for AdS$_4$ in terms of these Killing vector fields since they also are present for stationary asymptotically AdS$_4$ black hole space-times, the first one associated with symmetry under time-translation and the second one with rotational symmetry of the space-time \cite{Hawking:1972vc,Hollands:2006rj,Moncrief:2008mr}. The rotational Killng vector field $V_{(1)}$ has period $2\pi$ as required. We impose for any asymptotically AdS$_4$ space-time that $V_{(0)}$ for that space-time should asymptote to $V_{(0)}$ of AdS$_4$ as given by \eqref{ads4killing}. This ensures that the domain length \eqref{domlength} is well-defined for asymptotically AdS$_4$ space-times.

We proceed now to make a coordinate transformation in order to put the metric in the canonical form \eqref{genmetric}. However, since we shall consider below a more general class of metrics which includes \eqref{ads4} as special case, we perform the analysis for the general class of metrics here, and then specialize to AdS$_4$ afterwards. The general class of metrics is 
\begin{equation}
\label{genclass4}
ds^2 = - f (\rho) dt^2 + \frac{d\rho^2}{f(\rho)} + \rho^2 ( d\theta^2 + \sin^2 \theta d\phi^2 ) 
\end{equation}
for which we choose in general the commuting Killing vectors as \eqref{ads4killing}. Computing $r^2 = |\det G_{ij}|$ and making an ansatz for the $z$-coordinate, we write
\begin{equation}
\label{genans4}
r = \rho \sqrt{f(\rho)} \sin \theta \spa B(z) = A(\rho) \cos \theta
\end{equation}
Demanding now that $g^{rz}=0$ we find 
\begin{equation}
\label{aeq}
(\log A)' = \frac{2}{( \rho^2 f )'}
\end{equation}
while $B(z)$ remains undetermined. In particular this means that for any choice of the $B(z )$ function the metric is locally of the canonical form \eqref{genmetric}. Clearly the freedom of choosing $B(z )$ corresponds to the freedom in making coordinate transformations of the $z$ coordinate. We impose that $B'(z) > 0$. Using \eqref{aeq} we find 
\begin{equation}
\label{lamb4}
\lambda (r,z) = \frac{B'(z)}{A'  \big( \rho (r,z) \big) }
\end{equation}

We now specialize to the AdS$_4$ metric \eqref{ads4} with $f = 1 + \rho^2 / L^2$. Integrating \eqref{aeq} we find $A(\rho) =  c L \rho / \sqrt{L^2 + 2 \rho^2 }$ where $c$ is a constant of integration that we choose to absorb in the choice of the $B(z)$ function and instead put $c=1$. We thus find the following canonical coordinates $(r,z)$ for AdS$_4$
\begin{equation}
\label{ads4rz}
r = \rho \sqrt{f} \sin \theta \spa B(z) = \frac{L\rho}{\sqrt{L^2 + 2\rho^2 }} \cos \theta
\end{equation}
For $B(z)=z$ and in the limit $L/\rho \rightarrow \infty$ we notice that these $(r,z)$ coordinates reduce to the canonical coordinates for four-dimensional Minkowski-space (see for example \cite{Emparan:2001wk,Harmark:2004rm}).
The global patch of AdS$_4$ is covered by the $(r,z)$ coordinates \eqref{ads4rz} in the ranges
\begin{equation}
\label{ads4rzrange}
r \geq 0 \spa - \frac{L}{\sqrt{2}} < B(z) < \frac{L}{\sqrt{2}}
\end{equation}
Despite the fact that the range of $B(z)$ is bounded we shall see below that the $z$ interval, as measured using the invariant measure \eqref{domlength}, is infinite. In line with this we note that the points $B(z) = \pm L/\sqrt{2}$ are not included on the $z$-axis which means that the $z$-axis is topologically like $\R$, $i.e.$ the topology of an open interval.

For large $r$ we find from \eqref{lamb4} that $\lambda$ asymptotes to
\begin{equation}
\label{ads4lambda}
\lambda \rightarrow  B'(z) \left( \frac{ 2 r}{\sqrt{L^2-2B(z)^2} } \right)^{\frac{3}{2}}    \ \ \mbox{for} \ \ r\rightarrow \infty
\end{equation}
Thus, for any static/stationary asymptotically AdS$_4$ black hole space-time with a rotational Killing vector field the function $\lambda(r,z)$ in the canonical metric \eqref{genmetric} should be of this form for large $r$. 

\begin{figure}[h!]
\centerline{\includegraphics[scale=0.5]{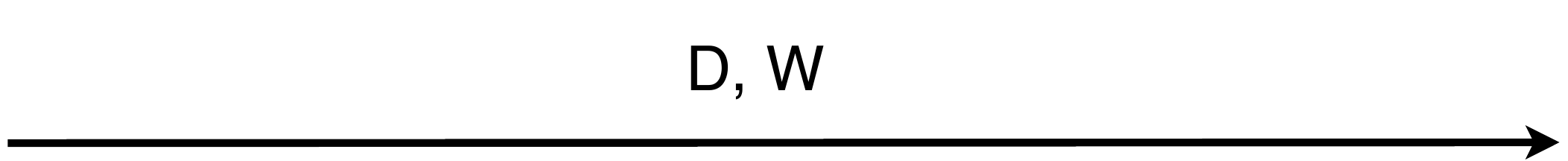}}
\vskip 0.2cm
\caption{\small Domain structure of AdS$_4$.}
\label{fig:ads4}
\begin{picture}(0,0)(0,0)
\put(204,37){\footnotesize $\phi$-axis}
\put(368,60){z}
\end{picture}
\end{figure}
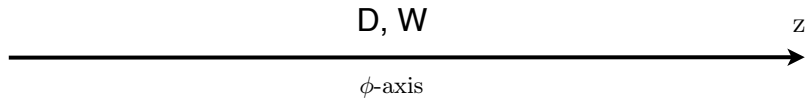

We now consider the domain structure of AdS$_4$. This is done by analyzing when $r=0$. In terms of the $(\rho,\theta)$ coordinates this happens when $\theta=0,\pi$. The domain structure is very simple: There is a single domain $D$ 
\begin{equation}
\begin{array}{c}
W = V_{(1)} \spa D = ( - \frac{1}{ \sqrt{2}}L , \frac{1}{ \sqrt{2}} L)
\end{array}
\end{equation}
We wrote here the interval in terms of the coordinate $\tilde{z} = B(z)$ (the actual values of the $z$ coordinate is not part of the domain structure, only the topological split up in domains plus the lengths of the domains). We note here that this is an open interval, thus $D$ is homeomorphic to $\R$. We illustrated the domain structure in Figure \ref{fig:ads4}.

As part of the domain structure we should also measure of the length of the domain $|D|$. Using \eqref{lamb4}
we compute
\begin{equation}
|D| = \int_{B^{-1} (-L/\sqrt{2})}^{B^{-1} (L/\sqrt{2})} \frac{B'(z)dz}{ \left( 1 - \frac{2B(z)^2}{L^2} \right)^{3/2} }  = \int_{-L/\sqrt{2}}^{L/\sqrt{2}} \frac{d\tilde{z}}{ \left( 1 - \frac{2\tilde{z}^2}{L^2} \right)^{3/2} }  = \infty
 \end{equation}
That $|D|= \infty$ means that it makes sense to state that $D= \R$ not only topologically but also geometrically. Thus the domain structure is the same as the one of four-dimensional Minkowski space-time \cite{Emparan:2001wk,Harmark:2004rm}.

\subsubsection*{Domain structure of AdS$_5$}

The metric of AdS$_5$ in the global patch can be written
\begin{equation}
\label{ads5}
ds^2 = - f dt^2 + \frac{d\rho^2}{f} + \rho^2 ( d\theta^2 + \sin^2 \theta d\phi_1^2 + \cos^2 \theta d\phi_2^2  ) \spa f(\rho) = 1 + \frac{\rho^2}{L^2}
\end{equation}
Here $L$ is the length-scale of AdS$_5$ in terms of which the cosmological constant is $\Lambda = - 6/ L^2$.
The ranges of the $(\rho,\theta)$ coordinates are $\rho \geq 0$ and $0\leq \theta \leq \pi/2$. In terms of the above metric AdS$_5$ has the three commuting Killing vector fields 
\begin{equation}
\label{ads5killing}
V_{(0)} = \frac{\partial}{\partial t} \spa V_{(1)} =  \frac{\partial}{\partial \phi_1}\spa V_{(2)} =  \frac{\partial}{\partial \phi_2}
\end{equation}
We choose to define the domain structure for AdS$_5$ in terms of these Killing vector fields since they are present for a large class of asymptotically AdS$_5$ space-times; the first one associated with symmetry under time-translation and the two others with rotational symmetry of the space-time. The rotational Killing vector fields $V_{(1)}$ and $V_{(2)}$ have periods $2\pi$ as required. We impose for any asymptotically AdS$_5$ space-time that $V_{(0)}$ for that space-time should asymptote to $V_{(0)}$ of AdS$_5$ as given by \eqref{ads5killing}. This ensures that the domain length \eqref{domlength} is well-defined for asymptotically AdS$_5$ space-times.

We proceed now with the analysis following closely the analysis of the AdS$_4$ case. We first need to find a coordinate transformation so that the metric is of the form \eqref{genmetric}. As for AdS$_4$, we consider a more general class of metrics which include \eqref{ads5} as special case. The general class of metrics is
\begin{equation}
\label{genclass5}
ds^2 = - f (\rho)dt^2 + \frac{d\rho^2}{f(\rho)} + \rho^2 ( d\theta^2 + \sin^2 \theta d\phi_1^2 + \cos^2 \theta d\phi_2^2  )
\end{equation}
for which we choose in general the commuting Killing vector as \eqref{ads5killing}. Computing $r^2 = |\det G_{ij} |$ and making an ansatz for the $z$-coordinate, we write
\begin{equation}
\label{genans5}
r = \frac{1}{2} \rho^2 \sqrt{f(\rho)} \sin 2\theta  \spa B(z) = A(\rho) \cos 2\theta 
\end{equation}
Demanding $g^{rz}=0$ we find
\begin{equation}
\label{aeqads5}
(\log A) ' = \frac{8\rho^2}{( \rho^4 f  )'}
\end{equation}
while $B(z)$ remains undetermined. As for AdS$_4$ the freedom in choosing $B(z)$ corresponds to the freedom in making coordinate transformations of the $z$ coordinate. We impose that $B'(z) > 0$. For $\lambda$ we find 
\begin{equation}
\label{lamb5}
\lambda(r,z) = \frac{\rho(r,z) B'(z)}{A'\big(\rho(r,z) \big) }
\end{equation}

We now specialize to the AdS$_5$ metric \eqref{ads5} with $f=1+\rho^2/L^2$. Integrating \eqref{aeqads5} we find the general solution $A(\rho) =  c L^2 \rho^2 / ( 2 L^2 + 3 \rho^2 )$ with $c$ a constant. Employing the freedom in choosing $B(z)$ we put $c=1$ without loss of generality. The canonical coordinates for AdS$_5$ are thus
\begin{equation}
\label{ads5rz}
r = \frac{1}{2} \rho^2 \sqrt{f} \sin 2\theta \spa B(z) = \frac{L^2 \rho^2}{2 L^2 + 3 \rho^2 } \cos 2\theta
\end{equation}
For $B(z)=z$ and in the limit $L/\rho \rightarrow \infty$ the $(r,z)$ coordinates becomes the canonical coordinates for five-dimensional Minkowski-space (see \cite{Emparan:2001wk,Harmark:2004rm}). 
The global patch of AdS$_5$ is covered by the $(r,z)$ coordinates \eqref{ads5rz} in the ranges
\begin{equation}
\label{ads5rzrange}
r \geq 0 \spa - \frac{L^2}{3} < B(z) < \frac{L^2}{3}
\end{equation}
As in the case of AdS$_4$ the range of $B(z)$ is bounded, however the $z$ axis does not include the points $B(z) = \pm L^2/3$ which suggests that the $z$-axis topologically is like $\R$. We remark on this further below when considering the domain structure.

Considering \eqref{lamb5} for large $r$ we find 
\begin{equation}
\label{ads5lambda}
\lambda \rightarrow \frac{9}{4} B'(z) \left( \frac{ 2 L r}{\sqrt{L^4-9B(z)^2} } \right)^{\frac{4}{3}}    \ \ \mbox{for} \ \ r\rightarrow \infty
\end{equation}
Thus, for any static/stationary asymptotically AdS$_5$ black hole space-time with two rotational Killing vectors field the function $\lambda(r,z)$ in the canonical metric \eqref{genmetric} should be of this form for large $r$. 

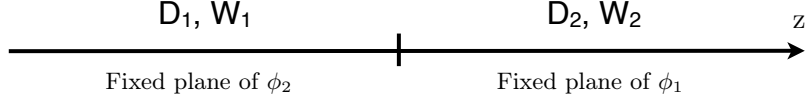
\begin{figure}[h!]
\centerline{\includegraphics[scale=0.5]{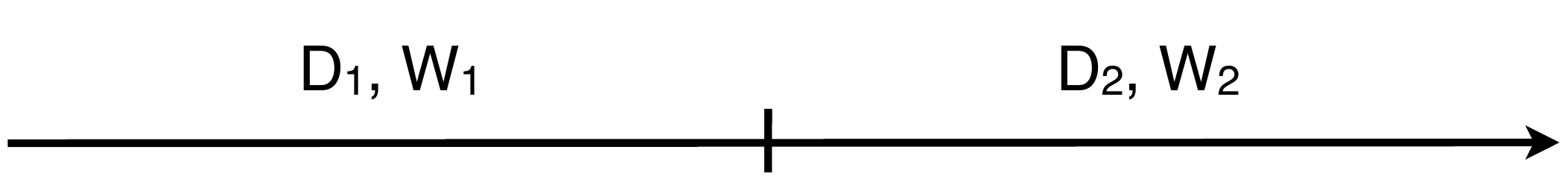}}
\vskip 0.2cm
\caption{\small Domain structure of AdS$_5$.}
\label{fig:ads5}
\begin{picture}(0,0)(0,0)
\put(108,40){\footnotesize Fixed plane of $\phi_2$}
\put(256,40){\footnotesize Fixed plane of $\phi_1$}
\put(368,63){z}
\end{picture}
\end{figure}

We now consider the domain structure of AdS$_5$. This is done by analyzing when $r=0$. In terms of the $(\rho,\theta)$ coordinates this happens when $\theta=0,\pi /2$. The domain structure consists of the two domains
\begin{equation}
\begin{array}{c}
W_1 = V_{(2)} \spa D_1 = ( - \frac{1}{3} L^2 , 0]
\\[3mm]
W_2 = V_{(1)} \spa D_2 = [ 0 , \frac{1}{3} L^2 )
\end{array}
\end{equation}
We wrote here the interval in terms of the coordinate $\tilde{z} = B(z)$ just as in the case of AdS$_4$ (again, the choice of $\tilde{z}$ is not part of the domain structure). We note here that the union of the two domains $D_1 \cup D_2$ is an open interval and thus is homeomorphic to $\R$. We illustrated the domain structure in Figure \ref{fig:ads5}.

As part of the domain structure we should also measure of the lengths of the domain $|D_1|$ and $|D_2|$. Using \eqref{lamb4}
we compute
\begin{equation}
|D_1| = \int_{B^{-1} (-L^2/3)}^{B^{-1} (0)} \frac{B'(z)dz}{ \left( 1 - \frac{3B(z)}{L^2} \right)^{2} }  = \int_{-L^2/3}^{0} \frac{d\tilde{z}}{ \left( 1 - \frac{3\tilde{z}}{L^2} \right)^{2} }  = \infty
 \end{equation}
One can similarly find $|D_2| = \infty$. Thus we can conclude that the union of the two domains $D_1 \cup D_2$, which we can think of as the $z$-axis, is both topologically and geometrically like $\R$, and the domain structure is therefore the same as the one of five-dimensional Minkowski space-time \cite{Emparan:2001wk,Harmark:2004rm}.

\subsection{Application to asymptotically AdS space-times}
\label{sec:appads}

We now turn to asymptotically AdS space-times in four and five dimensions. Since we require $p=D-2$ commuting Killing vector fields we can treat either four-dimensional stationary asymptotically AdS black hole space-times with one rotational Killing vector field, or five-dimensional stationary asymptotically AdS black hole space-times with two rotational Killing vector fields. However, we briefly consider higher-dimensional cases in Section \ref{sec:higherdim}. Note that the Rigidity theorems in the literature \cite{Hawking:1972vc,Hollands:2006rj,Moncrief:2008mr} also apply to asymptotically AdS space-times thus we are guaranteed the existence of at least one rotational Killing vector field.

\subsubsection*{Static spherically symmetric black holes in AdS$_4$}

We begin by considering a rather general class of static spherically symmetric asymptotically AdS$_4$ black hole space-times with metric of the form \eqref{genclass4} along with the following restrictions on the function $f(\rho)$ 
\begin{equation}
\label{staticBHads4}
 f (\rho) \simeq \frac{\rho^2}{L^2} \ \ \mbox{for}\ \  \rho \gg L  \spa \exists \rho_0 > 0 : f(\rho_0)=0 \ \ \mbox{and}\ \ f(\rho) > 0\ , \  ( \rho^2 f)' > 0 \ \ \mbox{for} \ \ \rho > \rho_0
\end{equation}
It follows from this that we have a static and spherically symmetric black hole space-time with an event horizon at $\rho=\rho_0$ of $S^2$ topology and which asymptotes to AdS$_4$ for $\rho/L \rightarrow \infty$. The ranges of the $(\rho,\theta)$ coordinates are $\rho \geq \rho_0$ and $0 \leq \theta \leq \pi$ for the part of the space-time outside the event horizon. Since the metric asymptote to \eqref{ads4} for $\rho/L \rightarrow \infty$ we should choose the two commuting Killing vectors as $V_{(0)} = \partial / \partial t$ and $V_{(1)} = \partial / \partial \phi$ as in \eqref{ads4killing}. This ensures that the domain length \eqref{domlength} is well-defined. 

We use the ansatz \eqref{genans4} for canonical $(r,z)$ coordinates for the metric \eqref{genclass4} with \eqref{staticBHads4}. The metric is in the canonical form \eqref{genmetric} provided $A(\rho)$ solves \eqref{aeq}. Note that since we demand $(\rho^2 f)' > 0$  it follows from \eqref{aeq} that $A(\rho)$ is positive and non-singular. The only restriction on $B(z)$ is $B'(z) > 0$. $\lambda(r,z)$ is given by \eqref{lamb4}. For $\rho \gg \rho_0, L$ we have $A(\rho) \simeq L/\sqrt{2}$ (making here a specific choice of the integration constant of Eq.~\eqref{aeq}). We thus have the same ranges \eqref{ads4rzrange} for the $(r,z)$ as for AdS$_4$.

\begin{figure}[h!]
\centerline{\includegraphics[scale=0.5]{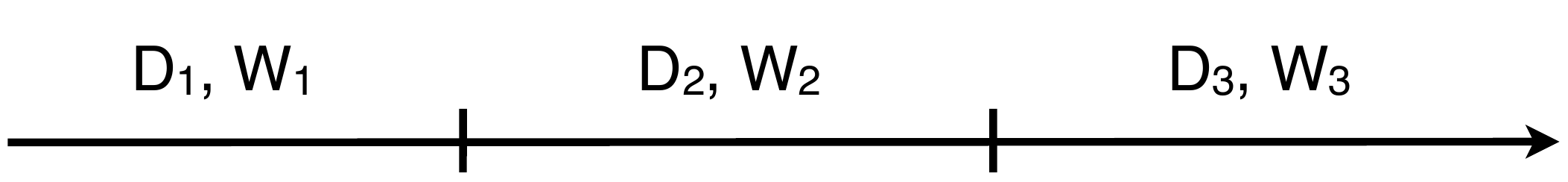}}
\vskip 0.2cm
\caption{\small Domain structure of black hole with spherical topology in AdS$_4$.}
\label{fig:staticBH4}
\begin{picture}(0,0)(0,0)
\put(100,40){\footnotesize $\phi$-axis}
\put(185,40){\footnotesize Event horizon}
\put(302,40){\footnotesize $\phi$-axis}
\put(368,63){z}
\end{picture}
\end{figure}
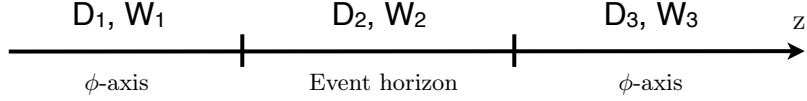

We can now read off the domain structure for the general class of static spherically symmetric asymptotically AdS$_4$ black hole space-times with metrics \eqref{genclass4} and \eqref{staticBHads4}. Using \eqref{genans4}  we see that $r=0$ when $\theta=0,\pi$ and $\rho=\rho_0$. We have three domains
\begin{equation}
\label{staticBH4}
\begin{array}{c}
W_1 = V_{(1)} \spa D_1 = ( -\frac{1}{\sqrt{2}} L , - \tilde{z}_0 ] \\[3mm]
W_2 = V_{(0)} \spa D_2 = [ - \tilde{z}_0 , \tilde{z}_0 ] \\[3mm]
W_3 = V_{(1)} \spa D_3 = [ \tilde{z}_0,  \frac{1}{\sqrt{2}} L ) 
\end{array}
\end{equation}
We wrote here the intervals in terms of the coordinate $\tilde{z} = B(z)$ (the choice of $\tilde{z}$ is not part of the domain structure). Here $\tilde{z}_0 = B(z_0) = A(\rho_0 )$. We illustrated the Domain structure \eqref{staticBH4} in Figure \ref{fig:staticBH4}.

We now measure the lengths of the three domains using Eq.~\eqref{domlength}. As we shall see the computations of the lengths are remarkably simple and do not require knowledge of the explicit form of $A(\rho)$. Consider first the domain $D_1$ corresponding to the axis of rotation. In the $(\rho,\theta)$ coordinates this is defined by $\theta=\pi$. Consider two $z$-values $z_1 < z_2$ within $D_1$. Then since from \eqref{genans4} we have $B' dz = - A' d\rho$ we find using \eqref{lamb4} that $\int_{z_1}^{z_2} \lambda dz = \rho_1 - \rho_2 $ where $(\rho,\theta)=(\rho_i,\pi)$ corresponds to $(r,z)=(0,z_i)$. Thus we find that $|D_1| = \infty$. The computation of $|D_3| = \infty$ proceeds in the same way. Consider instead the domain $D_2$ corresponding to the event horizon. In the $(\rho,\theta)$ coordinates this is defined by $\rho=\rho_0$. Using \eqref{genans4} we find $B(z)= A(\rho_0) \cos \theta$. Thus, we find
\begin{equation}
\label{staticBH4_D2}
|D_2| = \int_{-z_0}^{z_0} \lambda (0,z) dz = \frac{ \int_0^{\pi} \sin \theta d\theta} {(\log A)' |_{\rho= \rho_0} }  = \rho_0^2\,  f'(\rho_0)
\end{equation}
where we used \eqref{aeq} and \eqref{lamb4}.

It is important to notice that we found the domain structure above without the need to know what types of matter should be present for the metric \eqref{genclass4} and \eqref{staticBHads4} to be a solution of the Einstein equations in addition to a cosmological constant $\Lambda = - 3/ L^2$. $E.g.$ the black hole could be charged, coupled to a scalar field, have dust present outside the horizon, etc. A particular example could be the Reissner-Nordstr\" om-AdS$_4$ black hole which is a static charged spherically symmetric black hole that is a solution to 4D Einstein-Maxwell gravity with a cosmological constant $\Lambda = - 3/ L^2$. The metric in the solution is given by \eqref{genclass4} with
\begin{equation}
\label{RNA4}
f(\rho) = 1 + \frac{\rho^2}{L^2} -\frac{2\mu}{\rho} + \frac{q^2}{\rho^2}
\end{equation}
where $\mu$ is proportional to the mass and $q$ to the charge of the black hole (in general it is proportional to the square root of the sum of squares of the electric and the magnetic charge). This is obviously a special case of \eqref{staticBHads4} with $\rho_0$ being the largest real root of $\rho^2 f(\rho) = 0$. Hence from the above we get that
the domain structure is given by \eqref{staticBH4} with $|D_1|=|D_3|=\infty$ and 
\begin{equation}
\label{staticBH4_D2b}
|D_2| = 2\mu + \frac{2\rho_0^3}{L^2} - \frac{2q^2}{\rho_0}
\end{equation}
Note that for zero cosmological constant $\Lambda=0$ and zero charge $q=0$ this reduces to the domain/rod length computed for the 4D Schwarzschild black hole in \cite{Emparan:2001wk,Harmark:2004rm}.

\subsubsection*{Kerr-Newman-AdS$_4$ black hole}

The Kerr-Newman-AdS$_4$ black hole is a charged stationary asymptotically AdS$_4$ black hole with angular momentum and with an event horizon homeomorphic to a sphere. It is a solution to the 4D Einstein-Maxwell theory with a cosmological constant $\Lambda = - 3/ L^2$. The metric for the Kerr-Newman-AdS$_4$ black hole can be written as
\begin{equation}
 \label{kerrads4}
 \begin{array}{c} \ds
ds^{2}=-\frac{\Delta}{\Sigma}(dt-\frac{a}{\Xi}\sin^2\theta d\phi)^2 + \frac{\Delta_{\theta}\sin^2\theta}{\Sigma}(adt-\frac{\rho^2+a^2}{\Xi}d\phi)^2 + \Sigma \left[ \frac{d\rho^2}{\Delta} + \frac{d\theta^2}{\Delta_{\theta}} \right]
\\[4mm] \ds
\Delta (\rho) = (\rho^2+a^2)(1+\frac{\rho^2}{L^2}) - 2\mu\rho + q^2 \spa \Sigma (\rho,\theta) = \rho^2 + a^2\cos^2\theta
\\[4mm] \ds
 \Delta_{\theta} (\theta) = 1 - \frac{a^2}{L^2}\cos^2\theta \spa  \Xi=1-\frac{a^2}{L^2}
\end{array}
\end{equation}
The event horizon is located at $\rho=\rho_0$ with $\rho_{0}$ being the largest positive root of $\Delta(\rho)=0$. For $a=0$ we regain the Reissner-Nordstr\" om-AdS$_4$ black hole given by Eqs.~\eqref{genclass4} and \eqref{RNA4}. The parameter $a$ is the rotation parameter. A necessary condition for having a regular black hole space-time for $\rho \geq \rho_0$ is $a < L$.

In order to find the domain structure we should identify the choice of commuting Killing vector fields. These should be chosen such that we find that in the asymptotic region, where the metric asymptotes to \eqref{ads4}, the Killing vector fields asymptote to \eqref{ads4killing} chosen for AdS$_4$. For large $\rho$ we notice that the metric is not static in the sense that an observer at constant $\phi$ is rotating. We should thus transfer to a non-rotating frame%
\footnote{We included here a transformation of $\rho$ and $\theta$ which is needed in order to get the right form of the metric to compare with \eqref{ads4} asymptotically.}
\begin{equation}
\tilde{t} = t \spa \tilde{\phi} = \phi + \frac{a}{L^2} t \spa \Xi \, \tilde{\rho}^2 \sin^2 \tilde{\theta} = ( \rho^2 + a^2 ) \sin^2 \theta \spa \tilde{\rho}^2 \cos^2 \tilde{\theta} = \rho^2 \cos^2 \theta
\end{equation}
In this frame we find for large $\tilde{\rho}$
\begin{equation}
\label{asymptkerrads4}
ds^2 \simeq - \frac{\tilde{\rho}^2}{L^2}  d \tilde{t}^2 + \tilde{\rho}^2 \sin^2 \tilde{\theta}\, {d\tilde{\phi}}^2 + \frac{L^2}{\tilde{\rho}^2} d\tilde{\rho}^2 + \tilde{\rho}^2 d\tilde{\theta}^2
\end{equation} 
Thus, we should choose the Killing vector fields $V_{(0)} = \partial / \partial \tilde{t}$ and $V_{(1)} = \partial / \partial \tilde{\phi}$ which in terms of the coordinates used in \eqref{kerrads4} are
\begin{equation}
\label{kerrads4kill}
V_{(0)} =   \frac{\partial}{\partial t} - \frac{a}{L^2}  \frac{\partial}{\partial \phi} \spa V_{(1)} =  \frac{\partial}{\partial \phi}
\end{equation}
Indeed, $V_{(0)}$ is the time-translation Killing vector field in the asymptotic region and one should define the mass of the black hole with respect to this Killing vector field \cite{Gibbons:2004ai}. 
Note also that one can see directly from \eqref{asymptkerrads4} for $\tilde{\theta} \rightarrow 0$ that $V_{(1)}$ has period $2\pi$. 

From the metric \eqref{kerrads4} and the Killing vector fields \eqref{kerrads4kill} we compute $r^2= | \det G_{ij} |$ and we find the following ansatz for $z$
\begin{equation}
\label{kerrads4rz}
r = \frac{\sqrt{\Delta}}{\Xi} \sqrt{\Delta_\theta} \sin \theta \spa \tilde{B} (z) = A(\rho) \frac{\cos \theta}{\sqrt{1-\frac{a^2}{L^2} \cos 2\theta}}
\end{equation}
The ansatz for $z$ is found by demanding $g^{rz} = 0$ and assuming a factorized form. The coordinates $(r,z)$ give a metric on the canonical form \eqref{genmetric} provided $A(\rho)$ obeys
\begin{equation}
\label{aeqkerrads4}
( \log A )'  = \frac{2}{\Delta'} \left(  1 + \frac{a^2}{L^2} \right)
\end{equation}
We compute 
\begin{equation}
\label{lambkerrads4}
\lambda (r,z) = \frac{\sqrt{\Delta_\theta} | \frac{\partial r}{\partial \theta} |}{\sqrt{\Delta} | \frac{\partial z}{\partial \rho} |}  =\frac{\tilde{B}'(z)}{\Xi A'(\rho)} \left( 1 - \frac{a^2}{L^2} \cos 2\theta \right)^{\frac{3}{2}}
\end{equation}
It is not hard to see that for $\rho \rightarrow \infty$ the function $A(\rho)$ approaches a positive constant. We choose this constant to be $L \sqrt{\Xi}$ without loss of generality since we can freely choose the function $\tilde{B}(z)$. We thus have the coordinate ranges $r \geq 0$ and $| \tilde{B} (z) | < L$. It is possible to find the relation between the choice of $B(z)$ for AdS$_4$ in \eqref{ads4rz} and $\tilde{B}(z)$ by studying $\lambda(r,z)$ from \eqref{lambkerrads4} in the asymptotic region and comparing it to \eqref{ads4lambda}. However, it is not necessary to know these details for understanding the domain structure of Kerr-Newman-AdS$_4$.

We can now read off the domain structure for Kerr-Newman-AdS$_4$. Using \eqref{genans4}  we see that $r=0$ when $\theta=0,\pi$ and $\rho=\rho_0$. We have three domains
\begin{equation}
\label{domkerrads4}
\begin{array}{c}
W_1 = V_{(1)} \spa D_1 = ( - L , - \tilde{z}_0 ] \\[3mm]
W_2 = V_{(0)} + \Omega V_{(1)} \spa D_2 = [ - \tilde{z}_0 , \tilde{z}_0 ] \\[3mm]
W_3 = V_{(1)} \spa D_3 = [ \tilde{z}_0,  L ) 
\end{array}
\end{equation}
where
\begin{equation}
\Omega =  \frac{a ( \rho_0^2 + L^2 )}{L^2 ( \rho_0^2 + a^2 ) }
\end{equation}
is the angular velocity of the black hole \cite{Gibbons:2004ai} (with respect to a non-rotating frame at infinity).
We wrote here the intervals in terms of the coordinate $\tilde{z} = \tilde{B}(z)$ (the choice of $\tilde{z}$ is not part of the domain structure). Here $\tilde{z}_0 = \tilde{B}(z_0) = A(\rho_0 )/\sqrt{\Xi}$. We can illustrate the domain structure by Figure \ref{fig:staticBH4}.

We now measure the lengths of the three domains using Eq.~\eqref{domlength}. Consider first the domain $D_1$ corresponding to the axis of rotation and defined by $\theta=\pi$ in the $(\rho,\theta)$ coordinates. Consider two $z$-values $z_1 < z_2$ within $D_1$. From \eqref{kerrads4rz} and  \eqref{lambkerrads4} we have $\tilde{B}' dz = - A' d\rho / \sqrt{\Xi}$ and $\lambda = \tilde{B}' \sqrt{\Xi} / A'$ in $D_1$.  Using this we get $\int_{z_1}^{z_2} \lambda dz = \rho_1 - \rho_2$ where $(\rho,\theta)=(\rho_i,\pi)$ corresponds to $(r,z)=(0,z_i)$. Thus we find that $|D_1| = \infty$. The computation of $|D_3| = \infty$ proceeds in the same way. Consider instead the domain $D_2$ corresponding to the event horizon. In the $(\rho,\theta)$ coordinates this is defined by $\rho=\rho_0$. Using \eqref{kerrads4rz}, \eqref{aeqkerrads4} and \eqref{lambkerrads4} we find
\begin{equation}
\label{kerrads4_D2}
|D_2| = \int_{-z_0}^{z_0} \lambda (0,z) dz = \frac{\Delta'(\rho_0)}{2\Xi} \int_0^{\pi} \sin \theta d\theta  = \frac{\Delta'(\rho_0)}{\Xi} 
= \frac{2}{\Xi} \left( \mu + \frac{\rho_0^3}{L^2} - \frac{q^2 + a^2}{\rho_0}  \right)
\end{equation}
We see that in the static case $a=0$ this reduces to the domain length \eqref{staticBH4_D2b} of the event horizon of the Reissner-Nordstr\" om black hole . For zero cosmological constant $\Lambda=0$ and zero charge $q=0$ it reduces to the domain length of the event horizon of the Kerr black hole found in \cite{Harmark:2004rm}.%
\footnote{Note that it follows from the general considerations in Section \ref{sec:generaldom} that the domain length \eqref{kerrads4_D2} would have been the same if we had chosen $V_{(0)} = \partial / \partial t$ instead of \eqref{kerrads4kill}.}

\subsubsection*{Restrictions on horizon topologies in AdS$_4$}

More generally, we can consider what restrictions on the horizon topology our domain structure analysis imposes on asymptotically AdS$_4$ stationary black hole space-times. Since one cannot have two domains with a time-like direction next to each other in a regular space-time, a domain with time-like direction can only have domains with the space-like direction $V_{(1)}$ next to it. Assuming the domain structure space $B$ to be a connected set, this means that there are space-like domains with direction $V_{(1)}$ on each side of a domain with a time-like direction corresponding to a Killing horizon. From this we directly read off the topology of such an event horizon to be $S^2$.

\subsubsection*{Static spherically symmetric black holes in AdS$_5$}

We consider here a rather general class of static spherically symmetric asymptotically AdS$_5$ black hole space-times with metric of the form \eqref{genclass5} along with the following restrictions on the function $f(\rho)$ 
\begin{equation}
\label{staticBHads5}
 f (\rho) \simeq \frac{\rho^2}{L^2} \ \ \mbox{for}\ \  \rho \gg L  \spa \exists \rho_0 > 0 : f(\rho_0)=0 \ \ \mbox{and}\ \ f(\rho) > 0\ , \  ( \rho^4 f)' > 0 \ \ \mbox{for} \ \ \rho > \rho_0
\end{equation}
It follows from this that we have a static and spherically symmetric black hole space-time with an event horizon at $\rho=\rho_0$ of $S^3$ topology and which asymptotes to AdS$_5$ for $\rho/L \rightarrow \infty$. 
 The ranges of the $(\rho,\theta)$ coordinates are $\rho \geq \rho_0$ and $0 \leq \theta \leq \pi/2$ for the part of the space-time outside the event horizon. Since the metric asymptotes to \eqref{ads5} for $\rho/L \rightarrow \infty$ we should choose the three commuting Killing vectors as $V_{(0)} = \partial / \partial t$, $V_{(1)} = \partial / \partial \phi_1$ and $V_{(2)} = \partial / \partial \phi_2$ as in \eqref{ads5killing}. 

We use the ansatz \eqref{genans5} for canonical $(r,z)$ coordinates for the metric \eqref{genclass5} with \eqref{staticBHads5}. The metric is in the canonical form \eqref{genmetric} provided $A(\rho)$ solves \eqref{aeqads5}. Note that since we demand $(\rho^4 f)' > 0$  it follows that $A(\rho)$ is positive and non-singular. The only restriction on $B(z)$ is $B'(z) > 0$. $\lambda(r,z)$ is given by \eqref{lamb5}. For $\rho \gg \rho_0,L$ we have $A(\rho) \simeq L^2/3$ (making here a specific choice of the integration constant of Eq.~\eqref{aeqads5}). We thus have the same ranges \eqref{ads5rzrange} for the $(r,z)$ coordinates as for AdS$_5$.

\begin{figure}[h!]
\centerline{\includegraphics[scale=0.5]{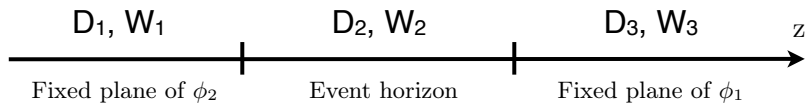}}
\vskip 0.2cm
\caption{\small Domain structure of black hole with $S^3$ topology in AdS$_5$.}
\label{fig:staticBH5}
\begin{picture}(0,0)(0,0)
\put(80,40){\footnotesize Fixed plane of $\phi_2$}
\put(185,40){\footnotesize Event horizon}
\put(279,40){\footnotesize Fixed plane of $\phi_1$}
\put(368,63){z}
\end{picture}
\end{figure}

We can now read off the domain structure for the general class of static spherically symmetric asymptotically AdS$_5$ black hole space-times with metrics \eqref{genclass5} and \eqref{staticBHads5}. Using \eqref{genans5}  we see that $r=0$ when $\theta=0,\pi/2$ and $\rho=\rho_0$. We have three domains
\begin{equation}
\label{staticBH5}
\begin{array}{c}
W_1 = V_{(2)} \spa D_1 = ( -\frac{1}{3} L^2 , - \tilde{z}_0 ] \\[3mm]
W_2 = V_{(0)} \spa D_2 = [ - \tilde{z}_0 , \tilde{z}_0 ] \\[3mm]
W_3 = V_{(1)} \spa D_3 = [ \tilde{z}_0,  \frac{1}{3} L^2 ) 
\end{array}
\end{equation}
We wrote here the intervals in terms of the coordinate $\tilde{z} = B(z)$ (the choice of $\tilde{z}$ is not part of the domain structure). Here $\tilde{z}_0 = B(z_0) = A(\rho_0 )$. We illustrated the Domain structure \eqref{staticBH5} in Figure \ref{fig:staticBH5}.

We now measure the lengths of the three domains using Eq.~\eqref{domlength}. Consider first the domain $D_1$. In the $(\rho,\theta)$ coordinates this is defined by $\theta=\pi/ 2$. Consider two $z$-values $z_1 < z_2$ within $D_1$. Then since from \eqref{genans5} we have $B' dz = - A' d\rho$ we find using \eqref{lamb5} that $\int_{z_1}^{z_2} \lambda dz = \frac{1}{2} ( \rho_1^2 - \rho_2^2)$ where $(\rho,\theta)=(\rho_i,\pi/2)$ corresponds to $(r,z)=(0,z_i)$. Thus we find that $|D_1| = \infty$. The computation of $|D_3| = \infty$ proceeds in the same way. Consider instead the domain $D_2$ corresponding to the event horizon. In the $(\rho,\theta)$ coordinates this is defined by $\rho=\rho_0$. Using \eqref{genans5} we find $B(z)= A(\rho_0) \cos 2 \theta$. Thus, we find
\begin{equation}
\label{staticBH5_D2}
|D_2| = \int_{-z_0}^{z_0} \lambda (0,z) dz = \frac{2 \rho_0 \int_0^{\pi/2} \sin 2\theta d\theta} {(\log A)' |_{\rho= \rho_0} }  = \frac{1}{4} \rho_0^3\,  f'(\rho_0)
\end{equation}
where we used \eqref{aeqads5} and \eqref{lamb5}.

It is important to notice that we found the domain structure above without the need to know what types of matter should be present for the metric \eqref{genclass5} and \eqref{staticBHads5} to be a solution of the Einstein equations in addition to a cosmological constant $\Lambda = - 6/ L^2$. A particular example that includes matter fields could be the Reissner-Nordstr\" om-AdS$_5$ black hole which is a static electrically charged spherically symmetric black hole that is a solution to 5D Einstein-Maxwell gravity with a cosmological constant $\Lambda = - 6/ L^2$. The metric in the solution is given by \eqref{genclass5} with
\begin{equation}
\label{RNA5}
f(\rho) = 1 + \frac{\rho^2}{L^2} -\frac{2\mu}{\rho^2} + \frac{q^2}{\rho^4}
\end{equation}
where $\mu$ is proportional to the mass and $q$ to the electric charge of the black hole. This is obviously a special case of \eqref{staticBHads5} with $\rho_0$ being the largest real root of $\rho^4 f(\rho) = 0$. Hence from the above we get that
the domain structure is given by \eqref{staticBH5} with $|D_1|=|D_3|=\infty$ and 
\begin{equation}
\label{staticBH5_D2b}
|D_2| = \mu + \frac{\rho_0^4}{2L^2} - \frac{q^2}{\rho_0^2}
\end{equation}
Note that for zero cosmological constant $\Lambda=0$ and zero charge $q=0$ this reduces to the domain/rod length computed for the 5D Schwarzschild-Tangherlini black hole in \cite{Emparan:2001wk,Harmark:2004rm}.

\subsubsection*{Kerr-AdS$_5$ black hole}

The Kerr-AdS$_5$ black hole is a stationary asymptotically AdS$_5$ black hole with angular momenta and with an event horizon homeomorphic to a three-sphere \cite{Hawking:1998kw,Gibbons:2004uw}. It is a solution of the Einstein equations with a cosmological constant $\Lambda = - 6 / L^2$. 
The metric can be written as
\begin{equation}
\label{kerrads5}
\begin{array}{rcl}
\ds ds^2 &=& \ds - \frac{\Delta}{\Sigma} \Big( dt - \sum_{i=1}^2 \frac{a_i \mu_i^2}{\Xi_i} d\phi_i \Big)^2 + \frac{\Delta_\theta}{\Sigma} \sum_{i=1}^2 \mu_i^2 \Big( a_i dt - \frac{\rho^2 + a_i^2}{\Xi_i} d\phi_i \Big)^2 \\[2mm] \ds && \ds + \frac{\rho^2 + L^2}{L^2 \rho^2 \Sigma} \Big( a_1 a_2 dt - \sum_{i=1}^2 \frac{a_i (\rho^2 + a_i^2 )}{\Xi_i} \mu_i^2 d\phi_i \Big)^2 + \Sigma \left[ \frac{d\rho^2}{\Delta} + \frac{d\theta^2}{\Delta_\theta} \right]
\end{array}
\end{equation}
with the definitions
\begin{equation}
\begin{array}{c} \ds
\Delta = \frac{1}{\rho^2} \Big(1 + \frac{\rho^2}{L^2} \Big) \prod_{i=1}^2 ( \rho^2 + a_i^2 ) - 2 \mu  \spa \Sigma = \rho^2 + \sum_{i=1}^2 a_i^2 (1- \mu_i^2)  \\[2mm]  \ds \Delta_\theta = 1 - \sum_{i=1}^2 \frac{a_i^2}{L^2} (1-\mu_i^2 ) \spa
\ds  \Xi_i = 1 - \frac{a_i^2}{L^2} \spa \mu_1 = \sin \theta \spa \mu_2 = \cos \theta 
\end{array}
\end{equation}
The event horizon is located at $\rho=\rho_0$ with $\rho_{0}$ being the largest positive root of $\rho^2 \Delta(\rho)=0$. For $a_1=a_2=0$ we regain the Schwarzschild-AdS$_5$ black hole given by Eqs.~\eqref{genclass5} and \eqref{RNA5} with $q=0$. The parameters $a_1$ and $a_2$ are the rotation parameters. Necessary conditions for having a regular black hole space-time for $\rho \geq \rho_0$ are $|a_i| < L$.

We now identify the choice of commuting Killing vector fields for studying the domain structure. 
As for the Kerr-Newman-AdS$_4$ black hole we should make a coordinate transformation such that we go to a non-rotating frame in the asymptotic region. This transformation is
\begin{equation}
\tilde{t} = t \spa \tilde{\phi}_i = \phi_i + \frac{a_i}{L^2} t \spa \Xi_1 \tilde{\rho}^2 \sin^2 \tilde{\theta} = (\rho^2 + a_1^2) \mu_1^2
\spa \Xi_2 \tilde{\rho}^2 \cos^2 \tilde{\theta} = (\rho^2 + a_2^2) \mu_2^2
\end{equation} 
With this transformation the metric has the same asymptotic behavior for $\tilde{\rho}\rightarrow \infty$ as \eqref{ads5} (identifying $(\tilde{\rho},\tilde{\theta})$ with $(\rho,\theta)$ in Eq.~\eqref{ads5}). 
Thus, we should choose the Killing vector fields as $V_{(0)} = \partial / \partial \tilde{t}$ and $V_{(i)} = \partial / \partial \tilde{\phi}_i$ which in terms of the coordinates used in \eqref{kerrads5} are
\begin{equation}
\label{kerrads5kill}
V_{(0)} = \frac{\partial}{\partial t} - \sum_{i=1}^2 \frac{a_i}{L^2} \frac{\partial}{\partial \phi_i} \spa V_{(1)} = \frac{\partial}{\partial \phi_1} \spa V_{(2)} = \frac{\partial}{\partial \phi_2} 
\end{equation}
$V_{(0)}$ is the time-translation Killing vector field in the asymptotic region and the mass of the black hole is defined with respect to this Killing vector field \cite{Gibbons:2004ai}. 

From \eqref{kerrads5} and \eqref{kerrads5kill} we find the following ansatz for $(r,z)$
\begin{equation}
\label{kerrads5rz}
r = \frac{\rho \sqrt{\Delta}}{2\Xi_1 \Xi_2} \sqrt{\Delta_\theta} \sin 2\theta \spa \tilde{B} (z) = A(\rho) g(\theta)
\end{equation}
The coordinates $(r,z)$ give a metric on the canonical form \eqref{genmetric} provided
\begin{equation}
(\log A)' = \frac{8 \rho^2 \sqrt{\epsilon^2 + \Xi_1 \Xi_2} }{( \rho^2 \Delta )'}
\spa
g(\theta) = \frac{4}{3 \epsilon}
\frac{3 \epsilon \cos 2\theta + \Xi_1 + \Xi_2 - 2\sqrt{ \epsilon^2 + \Xi_1 \Xi_2}  }{3 \epsilon \cos 2\theta + \Xi_1 + \Xi_2 + 2\sqrt{ \epsilon^2 +  \Xi_1 \Xi_2} } 
\end{equation}
with $\epsilon L^2 = a_2^2 - a_1^2$. Note that $g'(\theta) < 0$ for $0 < \theta < \pi/2$, $g'(0)=g'(\pi/2) = 0$ and that for $\epsilon \rightarrow 0$ we get $g(\theta) =  \cos (2 \theta ) / \Xi_1$.
We compute
\begin{equation}
\label{lambkerrads5}
\lambda (r,z) = \frac{\sqrt{\Delta_\theta} | \frac{\partial r}{\partial \theta} |}{\sqrt{\Delta} | \frac{\partial z}{\partial \rho} |}  = \frac{\rho \tilde{B}' (z) }{ A'(\rho) } \frac{\sqrt{\Delta_\theta} ( \sqrt{\Delta_\theta} \sin 2\theta )'}{2\Xi_1 \Xi_2 g(\theta)}
\end{equation}
For $\rho\rightarrow \infty$ $A(\rho)$ approaches a constant. We choose this as $L^2$ without loss of generality. With this we have the coordinate ranges $r \geq 0$ and $| \tilde{B} (z) | < ( \cdots ) L^2 $. We now read off the domain structure as
\begin{equation}
\label{domkerrads5}
\begin{array}{c}
W_1 = V_{(2)} \spa D_1 = \big( g(\frac{\pi}{2}) L^2 , g(\frac{\pi}{2}) A(\rho_0) \big] \\[3mm]
W_2 = V_{(0)} + \Omega_1 V_{(1)} + \Omega_2 V_{(2)}  \spa D_2 = \big[ g(\frac{\pi}{2}) A(\rho_0) , g(0) A(\rho_0) \big] \\[3mm]
W_3 = V_{(1)} \spa D_3 = \big[ g(0) A(\rho_0) ,  g(0) L^2 \big) 
\end{array}
\end{equation}
where the angular velocities measured relative to a non-rotating frame at infinity are  \cite{Gibbons:2004ai}
\begin{equation}
\Omega_i = \frac{a_i ( \rho_0^2 + L^2 )}{L^2 ( \rho_0^2 + a_i^2 ) }
\end{equation}
We can illustrate the domain structure by Figure \ref{fig:staticBH5}.
Concerning the length of the three domains \eqref{domkerrads5} we easily find $|D_1|=|D_3|=\infty$ and
\begin{equation}
| D_2 | = \frac{\rho_0 \Delta' (\rho_0) }{4 \Xi_1 \Xi_2 } = \frac{1}{\Xi_1 \Xi_2} \left[ \mu + \frac{\rho_0^4}{2L^2} - \frac{a_1^2+a_2^2}{2}- \frac{a_1^2a_2^2}{\rho_0^2}  -  \frac{a_1^2 a_2^2}{2L^2}    \right]
\end{equation}
which for $a_i=0$ reduces to \eqref{staticBH5_D2b} with $q=0$.

\subsubsection*{Black rings and multiple event horizons in AdS$_5$}

Above we have considered the domain structure of asymptotically AdS black holes using exact solutions. In all dimensions, the only known exact solutions for asymptotically AdS$_D$ black holes have a single event horizon homeomorphic to a $(D-2)$-sphere. This is in contrast with the situation for asymptotically flat black holes where exact solutions with new event horizon topologies have been found in five space-time dimensions. In particular, in addition to the Myers-Perry black holes \cite{Myers:1986un} with event horizon topology $S^3$, Emparan and Reall found an exact solution of a black ring with event horizon topology $S^1 \times S^2$, the $S^1$ stabilized by the centrifugal force in the plane of the $S^1$ \cite{Emparan:2001wn}. Furthermore, the Myers-Perry black holes and the black ring can be combined to give solutions with multiple disconnected event horizons \cite{Elvang:2007rd,Iguchi:2007is,Evslin:2007fv,Izumi:2007qx,Elvang:2007hs}. 

However, if we compare the domain structure of asymptotically AdS black holes to that of asymptotically flat black holes we see that it is identical for the same number of space-time dimensions and number of commuting Killing vector fields. It is therefore tempting to conjecture on this basis that the same domain structures of black holes should exist for asymptotically AdS as in the case of asymptotically flat black hole space-times. This means in particular that all the new topologies for event horizon found for five dimensional asymptotically flat black holes should also exist for five-dimensional asymptotically AdS black holes. A simple independent argument supporting this is that one can imagine taking an asymptotically flat black hole with maximal length scale $R$, in the sense that for distances much greater than $R$ away from the black hole the space-time is nearly flat. Then one could to a good approximation embed this in an asymptotically AdS space-time with cosmological constant length scale $L \gg R$. Thus, it seems likely that all the domain structures we find for asymptotically flat black holes space-times are also realized for asymptotically AdS black hole space-times. 

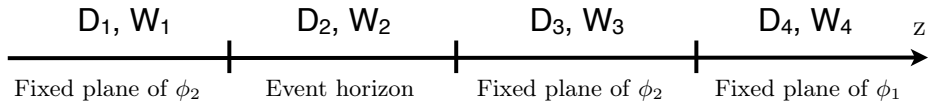
\begin{figure}[h!]
\centerline{\includegraphics[scale=0.5]{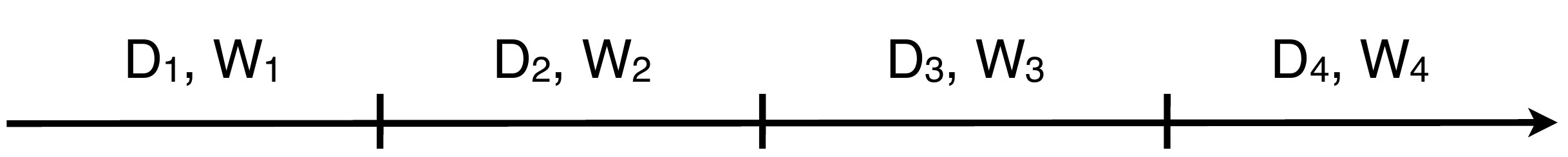}}
\vskip 0.2cm
\caption{\small Domain structure of black ring in AdS$_5$.}
\label{fig:adsblackring}
\begin{picture}(0,0)(0,0)
\put(50,40){\footnotesize Fixed plane of $\phi_2$}
\put(145,40){\footnotesize Event horizon}
\put(225,40){\footnotesize Fixed plane of $\phi_2$}
\put(315,40){\footnotesize Fixed plane of $\phi_1$}
\put(390,63){z}
\end{picture}
\end{figure}

A way to show the existence of additional types of asymptotically AdS black holes other than the ones for which we have exact solutions is by using the blackfold approach \cite{Emparan:2007wm}. In this way one can in particular find an approximate solution for a thin black ring in AdS$_5$ \cite{Caldarelli:2008pz,Armas:2010hz}. This would have the domain structure illustrated in Figure \ref{fig:adsblackring} with $W_{1} = W_3 = V_{(2)} = \partial / \partial \phi_2$, $W_{4} = V_{(1)} = \partial / \partial \phi_1$ and $W_2 = V_{(0)} + \Omega_1 V_{(1)} + \Omega_2 V_{(2)}$.
%

\subsubsection*{Restrictions on horizon topologies in AdS$_5$}

More generally, we can consider what restrictions on the horizon topology our domain structure analysis imposes on asymptotically AdS$_5$ stationary black hole space-times with two rotational Killing vector fields, as done in \cite{Hollands:2007aj} for five-dimensional asymptotically flat space-times. Consider again a domain with time-like direction, thus corresponding to a Killing horizon, and assume that the domain structure space $B$ is a connected set. The most general situation for the domain is that one has a domain $D'$ with space-like direction $V'$ on its left side and a domain $D''$ with space-like direction $V''$ on its right side. It follows from Theorem \ref{theo:dom} of Section \ref{sec:generaldom} that $V'$ and $V''$ should be of the form $V' = p' V_{(1)} + q' V_{(2)}$ and $V'' = p'' V_{(1)} + q'' V_{(2)}$ where $p'$ and $q'$ are mutually prime numbers, and the same for $p''$ and $q''$. It is straightforward to infer from this that the horizon topology is restricted to be either $S^3$, $S^2 \times S^1$ or a Lens-space.

\section{Domain structure of asymptotically dS space-times}
\label{sec:asymptds}

In this section we first consider the Canonical coordinates and the domain structure of dS$_4$ and dS$_5$. Subsequently we employ this to find the domain structure of asymptotically dS$_4$ and dS$_5$ stationary black hole space-times.

\subsection{Domain structure of dS}
\label{sec:domds}

\subsubsection*{General remarks on de Sitter space}

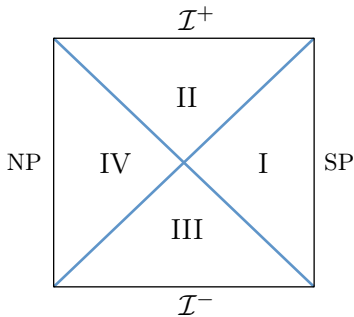
\begin{figure}[h!]
\vskip 0.3cm
\centerline{\includegraphics[scale=0.5]{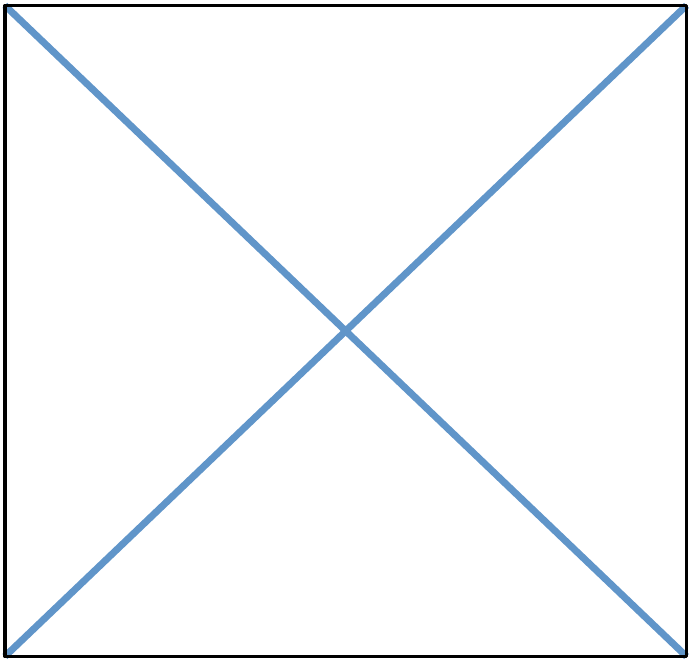}}
\caption{\small Penrose diagram of de Sitter space.}
\label{fig:desitter}
\begin{picture}(0,0)(0,0)
\put(250,85){I}
\put(219,110){II}
\put(217,60){III}
\put(190,85){IV}
\put(275,86){\footnotesize SP}
\put(155,86){\footnotesize NP}
\put(220,139){$\CI^+$}
\put(220,31){$\CI^-$}
\end{picture}
\end{figure}

In a particular choice of global coordinates the metric of $D$-dimensional de Sitter space dS$_D$ is
\begin{equation}
\label{globaldsmet}
ds^2= \frac{L^2}{\cos^2 T} ( - dT^2 + d\psi^2 + \sin^2 \psi \, d\Omega_{D-2}^2 )
\end{equation}
with the coordinate ranges $-\pi /2 < T < \pi / 2$ and $0 \leq \psi \leq \pi$. $D$-dimensional de Sitter space dS$_D$ is a solution to the Einstein equations with the only matter field being a cosmological constant  $\Lambda = (D-1)(D-2)/ (2 L^2)$. Using a conformally equivalent metric one finds the Penrose diagram for dS$_D$ depicted in Fig.~\ref{fig:desitter}. In the conformal completion $T = \pi /2 $ is future null infinity $\CI^+$ and $T = -\pi /2 $ is past null infinity $\CI^-$. Moreover, $\psi=0$ is the North pole (NP) and $\psi=\pi$ is the South pole (SP) of the $(D-1)$-dimensional sphere in the metric \eqref{globaldsmet}. The diagonal lines in Fig.~\ref{fig:desitter} are the cosmological event horizons for observers at the North and South poles. The causal future $\CO^+$ of an observer at the South pole is $\psi \geq \pi/2 - T$ (regions I and II in Fig.~\ref{fig:desitter}) while the causal past $\CO^-$ is $\psi \geq \pi/2 + T$ (regions I and III). The Southern causal diamond is the intersection $\CO^- \cap \CO^+$ (region I).

Our aim below is to describe the domain structure of de Sitter space and asymptotically de Sitter black hole space-times in the Southern causal diamond. The issue we address here is the choice of $V_{(0)}$ as the time-translation Killing vector field and how one imposes that this asymptotes to our choice for de Sitter space, $e.g.$ how do we define the notion of asymptotically de Sitter. 

Consider the following coordinates for the Southern causal diamond (region I) called static coordinates
\begin{equation}
\label{patch1}
\frac{\rho}{L} = \frac{\sin \psi}{\cos T} \spa \tanh \Big( \frac{t}{L} \Big) = - \frac{\sin T}{\cos \psi}
\end{equation}
This gives the metric 
\begin{equation}
\label{theds}
ds^2 = - f dt^2 + \frac{d\rho^2}{f} + \rho^2 d\Omega_{D-2}^2 \spa f = 1-\frac{\rho^2}{L^2}
\end{equation}
These coordinates only describe dS$_D$ in region I. In particular for $\rho = L$ we have the past and future cosmological event horizons. In this patch we choose
\begin{equation}
\label{dsV0}
V_{(0)} = \frac{\partial}{\partial t}
\end{equation}
To define the asymptotics of de Sitter space we need to consider $\rho \rightarrow \infty$. We define the asymptotics as going towards $\CI^+$ in region II. In region II we can analytically extend the metric \eqref{theds} by defining the $(t,\rho)$ coordinates as
\begin{equation}
\label{patch2}
\frac{\rho}{L} = \frac{\sin \psi}{\cos T} \spa \tanh \Big( \frac{t}{L} \Big) = - \frac{\cos \psi}{\sin T}
\end{equation}
Thus, we can extend our choice \eqref{dsV0} for $V_{(0)}$ into region II where our asymptotic region is. Note however that while in region I $V_{(0)}$ is the time-translation Killing vector field, in region II it is a space-like Killing vector field, as one can see from the metric \eqref{theds} for $\rho \geq L$. 

\subsubsection*{Domain structure of dS$_4$}

We consider now the domain structure of dS$_4$. In static coordinates the metric of dS$_4$ in the Southern causal diamond is (see also \eqref{theds})
\begin{equation}
\label{ds4}
ds^2 = - f dt^2 + \frac{d\rho^2}{f} + \rho^2 ( d\theta^2 + \sin^2 \theta d\phi^2 ) \spa f (\rho)= 1 - \frac{\rho^2}{L^2}
\end{equation}
Here $L$ is the length-scale of dS$_4$ in terms of which the cosmological constant is $\Lambda = 3/ L^2$. 
The ranges of the $(\rho,\theta)$ coordinates are $0 \leq \rho \leq L$ and $0\leq \theta \leq \pi$. 
In terms of the metric \eqref{ds4} dS$_4$ has the two commuting Killing vector fields 
\begin{equation}
\label{ds4killing}
V_{(0)} = \frac{\partial}{\partial t} \spa V_{(1)} =  \frac{\partial}{\partial \phi}
\end{equation}
We choose to define the domain structure for dS$_4$ in terms of these Killing vector fields since we expect these Killing vector fields are present for stationary asymptotically dS$_4$ black hole space-times, the first one associated with symmetry under time-translation and the second one with rotational symmetry of the space-time. The rotational Killng vector field $V_{(1)}$ has period $2\pi$ as required. 

Regarding our choice of $V_{(0)}$ in \eqref{ds4killing} it conforms with our choice for the $D$-dimensional de Sitter space as discussed above. In particular, we extend $V_{(0)}$ of \eqref{ds4killing} to the region $\rho \geq L$ beyond the future cosmological horizon such that it includes the asymptotic region near $\CI^+$. In line with this, we impose for a given asymptotically dS$_4$ space-time that in the analytically extended region beyond the future cosmological event horizon the $V_{(0)} $ chosen for that space-time should asymptote to our choice \eqref{ds4killing} when going towards $\CI^+$ in that space-time. This ensures that the domain length \eqref{domlength} is well-defined for asymptotically dS$_4$ space-times.

We proceed now to make a coordinate transformation in order to put the metric \eqref{ds4} in the canonical form \eqref{genmetric}. The metric \eqref{ds4} is in the general class \eqref{genclass4}. Thus, we use the ansatz \eqref{genans4} for the $(r,z)$ coordinates such that the metric is of the form  \eqref{genmetric} provided Eq.~\eqref{aeq} is fulfilled. Using the general solution to \eqref{aeq} we see that the $z$ coordinate is of the form
\begin{equation}
\label{undetzds4}
B(z) = \frac{\rho \cos \theta}{\sqrt{| L^2 - 2\rho^2 |}} 
\end{equation}
where we absorbed the choice of integration constant into the undetermined function $B(z)$. Note the solution \eqref{undetzds4} has two patches, one for $0 \leq \rho < L/\sqrt{2}$ and another for $L/\sqrt{2} < \rho \leq L$.

Considering \eqref{genans4} with $f(\rho) = 1-\rho^2/L^2$ we see that $r(\rho,\theta)$ has a maximum value $L/2$ attained at $\rho=L/\sqrt{2}$ and $\theta= \pi/2$. Thus, $0 \leq r \leq L/2$. $r$ is a function of $(\rho,\theta)$ in the ranges $0 \leq \rho \leq L$ and $0 \leq \theta \leq \pi$. A constant value $r < L/2$ corresponds to a closed curve in the $(\rho,\theta)$ plane which circles around the point $(\rho,\theta)=(L/\sqrt{2},\pi/2)$. Therefore, the topology of the $z$ coordinate is that of a periodic variable. Hence the function $B(z)$ in \eqref{undetzds4} should be in accordance with this. 

We consider now a specific choice of the function $B(z)$ with the right properties. We define the $z$ coordinate by
\begin{equation}
\label{zds4}
\frac{\sin ( \frac{z}{L})}{\sqrt{ | \cos ( \frac{z}{L}) | }} = \frac{\rho \cos \theta}{\sqrt{| L^2 - 2\rho^2 |}} 
\end{equation}
In detail $z$ is defined as a periodic variable of period $2\pi L$ and we define $z(\rho,\theta)$ such that the patch $- \pi L /2 < z < \pi L /2$ corresponds to the patch $0 \leq \rho < L/\sqrt{2}$ while the patch $\pi L /2 < z < 3\pi L /2$ corresponds to the patch $L/\sqrt{2} < \rho \leq L$. Concerning the smoothness of the $(r,z)$ coordinates for $r < L/2$ we see immediately that both the LHS and RHS of \eqref{zds4} are smooth functions in the two separate patches $- \pi L /2 < z < \pi L /2$ and $\pi L /2 < z < 3\pi L /2$. Close to $z = \pm \pi L/2$ we can instead write \eqref{zds4} as
\begin{equation}
\frac{\cos ( \frac{z}{L})}{\sin^2 (\frac{z}{L})} = \frac{L^2 - 2 \rho^2 }{\rho^2 \cos^2 \theta}
\end{equation}
from which we see that $z(\rho,\theta)$ is a smooth function in the neighborhoods of $z=-\pi L/2$ and $z=\pi L/2$. 
In the flat space limit $L \rightarrow \infty$ we regain $z$ as being on $\R$ and we find $z = \rho \cos \theta$ which is the standard choice of canonical coordinates of four-dimensional Minkowski space of \cite{Emparan:2001wk,Harmark:2004rm}.

Note that we for simplicity stick to a specific choice of the $z$ coordinate in the following but one can use any redefinition $\tilde{z} = \tilde{B} (z)$ of the above $z$ as long as $\tilde{B}$ is a smooth map from $S^1$ (with periodicity $2\pi L$) to $S^1$ with the same orientation. 
The Southern causal diamond of dS$_4$ is covered by the $(r,z)$ coordinates with $0 \leq r \leq L/2$ and with  $z$ periodic with period $2\pi L$. One can show that the $(r,z)$ coordinate system is not smooth at the point $r=L/2$.

\begin{figure}[h!]
\centerline{\includegraphics[scale=0.5]{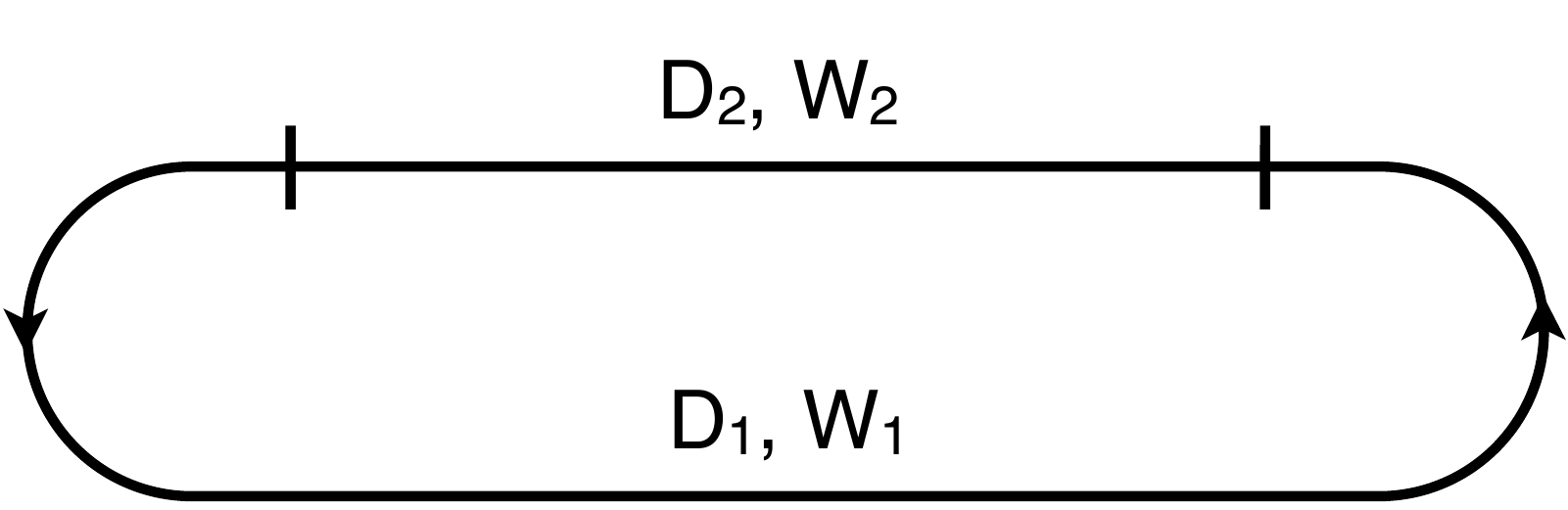}}
\vskip 0.2cm
\caption{\small Domain structure of dS$_4$.}
\label{fig:domds4}
\begin{picture}(0,0)(0,0)
\put(210,38){\footnotesize $\phi$-axis}
\put(180,86){\footnotesize Cosmological horizon}
\put(340,73){z}
\put(100,73){z}
\end{picture}
\end{figure}

We consider now the domain structure of dS$_4$ in the Southern causal diamond. This is done by analyzing when $r=0$. In terms of the $(\rho,\theta)$ coordinates $r=0$ when $\theta= 0, \pi$, corresponding to the rotational axis of symmetry, and $\rho=L$, corresponding to the cosmological event horizon. Define the angle $\alpha$ by $\cos \alpha = - \frac{1}{2} ( \sqrt{5}-1) $ and $0 < \alpha < \pi$. Then we can write the domain structure of dS$_4$ as
\begin{equation}
\begin{array}{c} \ds
W_1 = V_{(1)} \spa D_1 = [ -\alpha L , \alpha L ]\\[3mm] \ds
W_2 = V_{(0)} \spa D_2 = [ \alpha L , (2\pi -\alpha) L ]   
\end{array} 
\end{equation}
The domain $D_1$ is the axis of symmetry while the domain $D_2$ is the cosmological event horizon. The domain structure is illustrated in Figure \ref{fig:domds4}.
We now compute the domain lengths from \eqref{domlength}. Using the general expression \eqref{lamb4} for $\lambda(r,z)$ along with \eqref{genans4}, \eqref{aeq} and $f(\rho) = 1-\rho^2/L^2$ we find
\begin{equation}
\begin{array}{c} \ds
| D_1 | = \int_{- \alpha L}^{\alpha L} \lambda(0,z) dz = 2 \int_0^L d\rho = 2 L
\\[4mm] \ds
| D_2 | = \int_{\alpha L}^{(2\pi - \alpha) L} \lambda(0,z) dz = - \frac{1}{(\log A)' |_{\rho=L} } \int_0^\pi \sin \theta d\theta = - ( \rho^2 f )'  |_{\rho=L} =  2 L
\end{array}
\end{equation}
Notice that the computation did not require knowledge of the explicit choice of $z$ coordinate \eqref{zds4}. We see that the total invariant length of the $z$ axis is $4L$ as measured with \eqref{domlength}.

\subsubsection*{Domain structure of dS$_5$}

We proceed with the domain structure of dS$_5$. In static coordinates the metric of dS$_5$ in the Southern causal diamond is (see also \eqref{theds})
\begin{equation}
\label{ds5}
ds^2 = - f dt^2 + \frac{d\rho^2}{f} + \rho^2 ( d\theta^2 + \sin^2 \theta d\phi_1^2 + \cos^2 \theta d\phi_2^2 ) \spa f (\rho)= 1 - \frac{\rho^2}{L^2}
\end{equation}
Here $L$ is the length-scale of dS$_5$ in terms of which the cosmological constant is $\Lambda = 6/ L^2$. 
The ranges of the $(\rho,\theta)$ coordinates are $0 \leq \rho \leq L$ and $0\leq \theta \leq \pi/2$. 
In terms of the metric \eqref{ds5} dS$_5$ has the three commuting Killing vector fields 
\begin{equation}
\label{ds5killing}
V_{(0)} = \frac{\partial}{\partial t} \spa V_{(1)} =  \frac{\partial}{\partial \phi_1} \spa V_{(2)} =  \frac{\partial}{\partial \phi_2}
\end{equation}
We choose to define the domain structure for dS$_5$ in terms of these Killing vector fields since we expect that these Killing vector fields are present for stationary asymptotically dS$_5$ black hole space-times, the first one associated with symmetry under time-translation and the second one with rotational symmetry of the space-time. The rotational Killing vector fields $V_{(1)}$ and $V_{(2)}$ have periods $2\pi$ as required. 

Regarding our choice of $V_{(0)}$ in \eqref{ds5killing} it conforms with our choice for the $D$-dimensional de Sitter space as discussed above, as also discussed for dS$_4$. We extend $V_{(0)}$ of \eqref{ds5killing} to the region $\rho \geq L$ beyond the future cosmological horizon such that it includes the asymptotic region near $\CI^+$ and impose for a given asymptotically dS$_5$ space-time that in the analytically extended region beyond the future cosmological event horizon the $V_{(0)} $ chosen for that space-time should asymptote to our choice \eqref{ds5killing} when going towards $\CI^+$ in that space-time. This ensures that the domain length \eqref{domlength} is well-defined for asymptotically dS$_5$ space-times.

We now find coordinates which transform the metric \eqref{ds5} to the canonical form \eqref{genmetric}. The metric \eqref{ds5} is in the general class \eqref{genclass5}. Thus, we use the ansatz \eqref{genans5} for the $(r,z)$ coordinates such that the metric is of the form  \eqref{genmetric} provided Eq.~\eqref{aeqads5} is fulfilled. Using the general solution to \eqref{aeqads5} we can write
\begin{equation}
B(z) = \frac{\rho^2 \cos 2\theta}{2 L^2 - 3\rho^2} 
\end{equation}
where we absorbed the choice of integration constant into the function $B(z)$. Clearly this solution has two patches, one for $0 \leq \rho < \sqrt{2}L/\sqrt{3}$ and another for $\sqrt{2}L/\sqrt{3} < \rho \leq L$.
Considering \eqref{genans5} with $f(\rho) = 1-\rho^2/L^2$ we see that $r(\rho,\theta)$ has a maximum value $\sqrt{3}L^2/9$ attained at $\rho= \sqrt{2} L / \sqrt{3}$ and $\theta= \pi/4$. As for dS$_4$ this means that the curves of constant $r$ in the $(r,\theta)$ plane are closed curves when $r < \sqrt{3}L^2/9$. Therefore, the topology of the $z$ coordinate is that of a periodic variable.

We consider now a specific choice of the function $B(z)$ with the right properties. We define the $z$ coordinate by
\begin{equation}
\label{zds5}
\tan \Big( \frac{z}{L^2} \Big) = \frac{\rho^2 \cos 2\theta}{2 L^2 - 3\rho^2} 
\end{equation}
In detail $z$ is defined as a periodic variable of period $2\pi L^2$ and we define $z(\rho,\theta)$ such that the patch $- \pi L^2 /2 < z < \pi L^2 /2$ corresponds to the patch $0 \leq \rho < \sqrt{2}L/\sqrt{3}$ while the patch $\pi L^2 /2 < z < 3\pi L^2 /2$ corresponds to the patch
$\sqrt{2}L/\sqrt{3} < \rho \leq L$.
Concerning the smoothness of the $(r,z)$ coordinates for $r < \sqrt{3}L^2/9$ we see immediately that both the LHS and RHS of \eqref{zds5} are smooth functions in the two separate patches $- \pi L^2 /2 < z < \pi L^2 /2$ and $\pi L^2 /2 < z < 3\pi L^2 /2$. Close to $z = \pm \pi L^2/2$ we can instead write \eqref{zds5} as
\begin{equation}
\cot \Big( \frac{z}{L^2} \Big) = \frac{2 L^2 - 3\rho^2}{\rho^2 \cos 2\theta} 
\end{equation}
from which we see that $z(\rho,\theta)$ is a smooth function in the neighborhoods of $z=-\pi L^2/2$ and $z=\pi L^2/2$. 
In the flat space limit $L \rightarrow \infty$ we regain $z$ as being on $\R$ and we find $z = \frac{1}{2} \rho^2 \cos 2 \theta$ which is the standard choice of canonical coordinates of five-dimensional Minkowski space of \cite{Emparan:2001wk,Harmark:2004rm}.

For simplicity we stick to a specific choice of the $z$ coordinate in the following but one can use any redefinition $\tilde{z} = \tilde{B} (z)$ of the above $z$ as long as $\tilde{B}$ is a smooth map from $S^1$ (with periodicity $2\pi L^2$) to $S^1$ with the same orientation. 
The Southern causal diamond of dS$_5$ is covered by the $(r,z)$ coordinates with $0 \leq r \leq \sqrt{3}L^2/9$ and with  $z$ periodic with period $2\pi L^2$. Note that the $(r,z)$ coordinate system chosen above is not smooth at the point $r=\sqrt{3} L^2 / 9$.

\begin{figure}[h!]
\centerline{\includegraphics[scale=0.5]{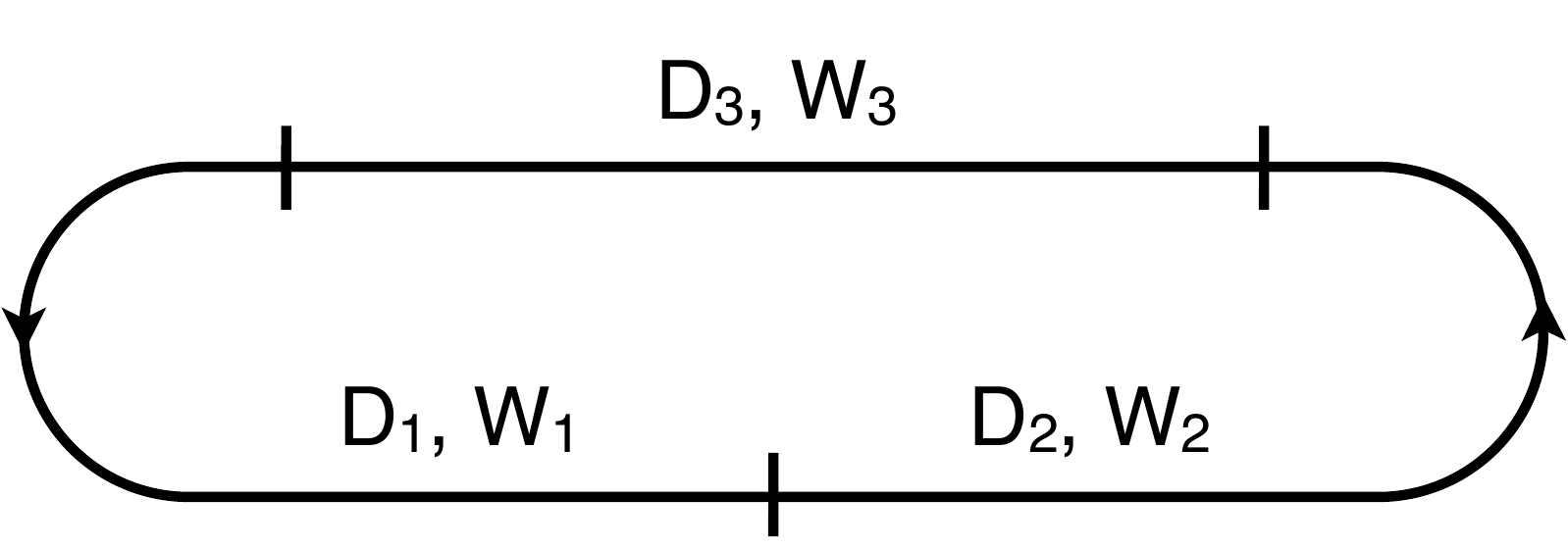}}
\vskip 0.2cm
\caption{\small Domain structure of dS$_5$.}
\label{fig:domds5}
\begin{picture}(0,0)(0,0)
\put(136,42){\footnotesize Fixed plane of $\phi_2$}
\put(236,42){\footnotesize Fixed plane of $\phi_1$}
\put(180,91){\footnotesize Cosmological horizon}
\put(340,78){z}
\put(100,78){z}
\end{picture}
\end{figure}

We consider now the domain structure of dS$_5$ in the Southern causal diamond. This is done by analyzing when $r=0$. In terms of the $(\rho,\theta)$ coordinates $r=0$ when $\theta= 0, \pi /2$, corresponding to the two rotational planes of symmetry, and $\rho=L$, corresponding to the cosmological event horizon. The domain structure is
\begin{equation}
\begin{array}{c} \ds
W_1 = V_{(2)} \spa D_1 = \Big[ -\frac{3}{4} \pi L^2 , 0 \Big]\\[3mm] \ds
W_2 = V_{(1)} \spa D_2 = \Big[  0 , \frac{3}{4} \pi L^2 \Big]\\[3mm] \ds
W_3 = V_{(0)} \spa D_3 = \Big[ \frac{3}{4} \pi L^2 , \frac{5}{4} \pi L^2 \Big]   
\end{array} 
\end{equation}
The domains $D_1$ and $D_2$ are the fixed planes for the $\phi_2$ and $\phi_1$ rotation angles, respectively, while $D_2$ is the cosmological event horizon.
The domain structure is illustrated in Figure \ref{fig:domds5}.
We now compute the domain lengths from \eqref{domlength}. Using the general expression \eqref{lamb5} for $\lambda(r,z)$ along with \eqref{genans5}, \eqref{aeqads5} and $f(\rho) = 1-\rho^2/L^2$ we find
\begin{equation}
\begin{array}{c} \ds
| D_1 | = \int_{- \frac{3}{4} \pi L}^{0} \lambda(0,z) dz = \int_0^L \rho d\rho =  \frac{1}{2} L^2
\\[4mm] \ds
| D_2 | = \int_{0}^{ \frac{3}{4} \pi L} \lambda(0,z) dz = \int_0^L \rho d\rho =  \frac{1}{2} L^2
\\[4mm] \ds
| D_3 | = \int_{\frac{3}{4} \pi L}^{\frac{5}{4} \pi L} \lambda(0,z) dz = - \frac{2L}{(\log A)' |_{\rho=L} } \int_0^{\pi /2} \sin 2 \theta \, d\theta = - \frac{1}{4L} ( \rho^4 f )'  |_{\rho=L} =  \frac{1}{2} L^2
\end{array}
\end{equation}
Notice that the computation did not require knowledge of the explicit choice of $z$ coordinate \eqref{zds5}. We see that the total invariant length of the $z$ axis is $3L^2 / 2$ as measured with \eqref{domlength}.

\subsection{Application to asymptotically dS space-times}
\label{sec:appds}

We now turn to asymptotically dS space-times in four and five dimensions. Since we require $p=D-2$ commuting Killing vector fields we can treat either four-dimensional stationary asymptotically dS black holes space-times with one rotational Killing vector field, or five-dimensional stationary asymptotically dS black holes space-times with two rotational Killing vector fields. However, we briefly consider higher-dimensional cases in Section \ref{sec:higherdim}. 

\subsubsection*{Static spherically symmetric black holes in dS$_4$}

We consider first a rather general class of static spherically symmetric asymptotically dS$_4$ black hole space-times with metric of the form \eqref{genclass4} along with the following restrictions on the function $f(\rho)$ 
\begin{equation}
\label{staticBHds4}
\begin{array}{c} \ds
 \exists \rho_0, \rho_s, \rho_c:  \rho_c> \rho_s > \rho_0  > 0 \ , \ f(\rho_0)=f(\rho_c) =0 \ , \  ( \rho^2 f)' |_{\rho=\rho_s} = 0 \ ,   \\[3mm]
 (\rho^2 f)' > 0 \ \mbox{for} \ \rho_0 \leq \rho < \rho_s \ , \  (\rho^2 f)' < 0 \ \mbox{for} \ \rho_s < \rho \leq \rho_c \ ,
 \\[3mm] \ds  f (\rho) \simeq -\frac{\rho^2}{L^2} \ \ \mbox{for}\ \  \rho \gg \rho_c  
\end{array}
\end{equation}
It follows from this that we have a static and spherically symmetric black hole space-time with an event horizon at $\rho=\rho_0$ of $S^2$ topology and with a cosmological horizon at $\rho=\rho_c$ also with $S^2$ topology. The region $\rho_0 \leq \rho \leq \rho_c$ is the causal diamond region for which we would like to find the domain structure. Instead in the region $\rho \geq \rho_c$ we find the future null infinity $\CI^+$ for $\rho/L \rightarrow \infty$. 
Since the metric asymptote to \eqref{ds4} for $\rho/L \rightarrow \infty$ we should choose the two commuting Killing vectors as $V_{(0)} = \partial / \partial t$ and $V_{(1)} = \partial / \partial \phi$ as in \eqref{ds4killing}. This ensures that the domain length \eqref{domlength} is well-defined. 

We use the ansatz \eqref{genans4} for canonical $(r,z)$ coordinates for the metric \eqref{genclass4} with \eqref{staticBHds4}. The metric is in the canonical form \eqref{genmetric} provided $A(\rho)$ solves \eqref{aeq}. Note that the requirements \eqref{staticBHds4} ensures that the function $\rho^2 f(\rho)$ only has a single extremum for $\rho_0 < \rho < \rho_c$ which is the maximum at $\rho=\rho_s$. We write the expansion around $\rho=\rho_s$ as $\rho^2 f(\rho) = \beta^2 - \alpha (\rho-\rho_s)^2 + \CO( (\rho-\rho_s)^3)$ with $\alpha, \beta > 0$. With this one finds 
\begin{equation}
\label{thea_dsbh4}
A(\rho) \simeq c  \left| 1 - \frac{\rho}{\rho_s}  \right|^{-\frac{1}{\alpha}} \ \ \mbox{for}\ \ \rho \simeq \rho_s
\end{equation}
Thus, just as for pure dS$_4$ we have two patches for the $(r,z)$ coordinate system for the causal diamond defined by $\rho_0 \leq \rho \leq \rho_s$ and $\rho_s \leq \rho \leq \rho_c$. We also note that $0 \leq r \leq \beta$ with the maximal value of $r$ reached in the point $(\rho,\theta)=(\rho_s,\pi/2)$. The topology of the $z$ coordinate is that of a periodic coordinate. We choose the period as $2\pi L$ and we make the following choice of $B(z)$ consistent with \eqref{thea_dsbh4} 
\begin{equation}
\label{theb_dsbh4}
\sin (\frac{z}{L}) \left| \cos ( \frac{z}{L} ) \right|^{-\frac{1}{\alpha}} = A(\rho) \cos \theta
\end{equation}
Just as for pure dS$_4$, this ensures a smooth function $z(\rho,\theta)$ for $z \neq \pm \pi L /2$ and $r < \beta$. For $z \simeq \pm \pi L /2$ we find $\cos (z/L) /  |\sin (z/L) |^\alpha \simeq c^{-\alpha} ( 1 - \rho / \rho_s) $ which ensures smoothness around $z = \pm \pi L /2$.

\begin{figure}[h!]
\centerline{\includegraphics[scale=0.5]{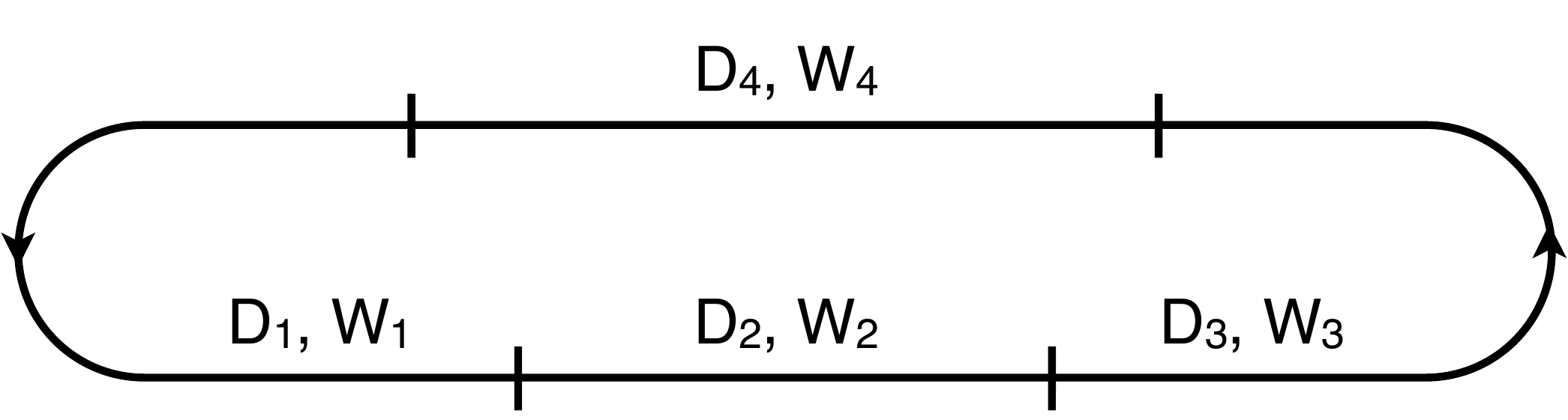}}
\vskip 0.2cm
\caption{\small Domain structure of black hole with spherical topology in dS$_4$.}
\label{fig:domdsbh4}
\begin{picture}(0,0)(0,0)
\put(120,42){\footnotesize $\phi$-axis}
\put(195,42){\footnotesize Event horizon}
\put(298,42){\footnotesize $\phi$-axis}
\put(180,91){\footnotesize Cosmological horizon}
\put(376,78){z}
\put(64,78){z}
\end{picture}
\end{figure}
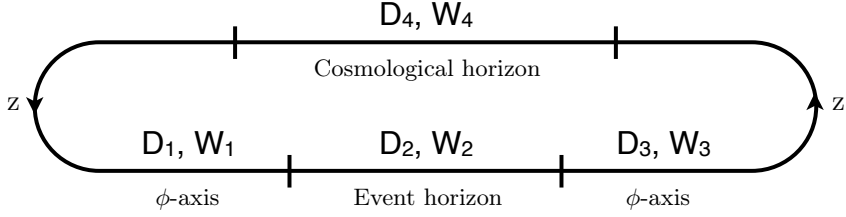

We can now read off the domain structure for the general class of static spherically symmetric asymptotically dS$_4$ black hole space-times with metrics \eqref{genclass4} and \eqref{staticBHds4}. Using \eqref{genans4}  we see that $r=0$ when $\theta=0,\pi$ and $\rho=\rho_0,\rho_c$. We have therefore the four domains
\begin{equation}
\label{domdsbh4}
\begin{array}{c}
W_1 = V_{(1)} \spa D_1 = [ - z_c , - z_0 ] \\[3mm]
W_2 = V_{(0)} \spa D_2 = [ - z_0 , z_0 ] \\[3mm]
W_3 = V_{(1)} \spa D_3 = [ z_0,  z_c ] \\[3mm]
W_4 = V_{(0)} \spa D_4 = [ z_c,  2\pi L - z_c ] 
\end{array}
\end{equation}
where we defined $z_0$ and $z_c$ by $0 < z_0/L < \pi / 2 < z_c/L < \pi $, $B(z_0) = A(\rho_0)$ and $B(z_c) = A(\rho_c)$.
The domains $D_1$ and $D_3$ are parts of the axis of symmetry for the $\phi$ angle while $D_2$ and $D_4$ are the black hole event horizon and the cosmological horizon, respectively.
The domain structure \eqref{domdsbh4} is illustrated in Figure \ref{fig:domdsbh4}.
Using \eqref{genans4}, \eqref{aeq} and \eqref{lamb4} we easily calculate the invariant domain lengths defined by Eq.~\eqref{domlength}
\begin{equation}
|D_1| = |D_3| = \rho_c - \rho_0 \spa |D_2| = \rho_0^2 f'(\rho_0) \spa |D_4| = - \rho_c^2 f'(\rho_c)
\end{equation}
The total invariant length is thus $2\rho_c - 2\rho_0 + \rho_0^2 f'(\rho_0) - \rho_c^2 f'(\rho_c)$.

Just as in the AdS case, it is important to remark that we found the domain structure above without the need to know what types of matter that should be present for the metric \eqref{genclass4} and \eqref{staticBHds4} to be a solution of the Einstein equations in addition to a cosmological constant $\Lambda = 3/ L^2$. A particularly example is the Reissner-Nordstr\" om-dS$_4$ black hole which is a static charged spherically symmetric black hole that is a solution to 4D Einstein-Maxwell gravity with a cosmological constant $\Lambda = 3/ L^2$. The metric in the solution is given by \eqref{genclass4} with
\begin{equation}
\label{RNdS4}
f(\rho) = 1 - \frac{\rho^2}{L^2} -\frac{2\mu}{\rho} + \frac{q^2}{\rho^2}
\end{equation}
where $\mu$ is proportional to the mass and $q$ to the charge of the black hole (in general it is proportional to the square root of the sum of squares of the electric and the magnetic charge). Consider now a choice of parameters $L$, $\mu$ and $q$ for which the requirements \eqref{staticBHds4} are met, with $\rho_0$ and $\rho_c$ marking the event horizon and the cosmological, respectively.
From the above we get then that
the domain structure is given by \eqref{domdsbh4} with 
\begin{equation}
\label{RNdS4_D2b}
|D_1| = |D_3| = \rho_c - \rho_0 \spa 
|D_2| = 2\mu - \frac{2\rho_0^3}{L^2} - \frac{2q^2}{\rho_0} \spa 
|D_4| =  \frac{2\rho_c^3}{L^2} - 2\mu  + \frac{2q^2}{\rho_c}
\end{equation}
%

\subsubsection*{Kerr-Newman-dS$_4$ black hole}

The Kerr-Newman-dS$_4$ black hole is a charged stationary asymptotically dS$_4$ black hole with angular momentum and with an event horizon homeomorphic to a sphere. It is a solution to the 4D Einstein-Maxwell theory with a cosmological constant $\Lambda = 3/ L^2$. The metric for the Kerr-Newman-dS$_4$ black hole can be written as
\begin{equation}
 \label{kerrds4}
 \begin{array}{c} \ds
ds^{2}=-\frac{\Delta}{\Sigma}(dt-\frac{a}{\Xi}\sin^2\theta d\phi)^2 + \frac{\Delta_{\theta}\sin^2\theta}{\Sigma}(adt-\frac{\rho^2+a^2}{\Xi}d\phi)^2 + \Sigma \left[ \frac{d\rho^2}{\Delta} + \frac{d\theta^2}{\Delta_{\theta}} \right]
\\[4mm] \ds
\Delta (\rho) = (\rho^2+a^2)(1-\frac{\rho^2}{L^2}) - 2\mu\rho + q^2 \spa \Sigma (\rho,\theta) = \rho^2 + a^2\cos^2\theta
\\[4mm] \ds
 \Delta_{\theta} (\theta) = 1 + \frac{a^2}{L^2}\cos^2\theta \spa  \Xi=1+\frac{a^2}{L^2}
\end{array}
\end{equation}
The parameter $a$ is the rotation parameter. For $a=0$ we regain the Reissner-Nordstr\" om-dS$_4$ black hole given by Eqs.~\eqref{genclass4} and \eqref{RNdS4}. The metric \eqref{kerrds4} can be obtained from the metric \eqref{kerrads4} of the Kerr-Newman-AdS$_4$ black hole by analytic continuation $L^2 \rightarrow - L^2$.
We assume the parameters $L$, $\mu$, $q$ and $a$ are such that 
\begin{equation}
\begin{array}{c} \ds
 \exists \rho_0, \rho_s, \rho_c:  \rho_c> \rho_s > \rho_0  > 0 \ , \ \Delta(\rho_0)=\Delta(\rho_c) = \Delta'(\rho_s) =0  \ ,   \\[3mm]
 \Delta'(\rho) > 0 \ \mbox{for} \ \rho_0 \leq \rho < \rho_s \ , \  \Delta'(\rho) < 0 \ \mbox{for} \ \rho_s < \rho \leq \rho_c 
 \end{array}
\end{equation}
At $\rho=\rho_0$ we have an event horizon with $S^2$ topology and at $\rho=\rho_c$ a cosmological horizon also with $S^2$ topology. The causal diamond within which we consider the domain structure is defined by $\rho_0 \leq \rho \leq \rho_c$. The metric \eqref{kerrds4} can be analytically extended for $\rho \geq \rho_c$ and we find the asymptotic region for $\rho \gg \rho_c$ near future null infinity $\CI^+$. 

Finding the domain structure basically consists in combining the analysis of static spherically symmetric dS$_4$ black holes \eqref{staticBHds4} with the analytical continuation $L^2 \rightarrow - L^2$ of the analysis of the Kerr-Newman-AdS$_4$ black hole \eqref{kerrads4}. We will therefore be brief in the following. 

The correct choice of the Killing vector fields is
\begin{equation}
\label{kerrds4kill}
V_{(0)} =   \frac{\partial}{\partial t} + \frac{a}{L^2}  \frac{\partial}{\partial \phi} \spa V_{(1)} =  \frac{\partial}{\partial \phi}
\end{equation}
since this ensures that $V_{(0)}$ asymptotes to the Killing vector field $\partial / \partial t$ for the metric \eqref{ds4} of dS$_4$ for $\rho\rightarrow \infty$, $i.e.$ near null infinity $\CI^+$. The coordinates $(r,z)$ gives a metric on the canonical form \eqref{genmetric} with
\begin{equation}
\label{kerrds4rz}
r = \frac{\sqrt{\Delta}}{\Xi} \sqrt{\Delta_\theta} \sin \theta \spa \tilde{B} (z) = A(\rho) \frac{\cos \theta}{\sqrt{1+\frac{a^2}{L^2} \cos 2\theta}}
\end{equation}
provided $A(\rho)$ obeys
\begin{equation}
\label{aeqkerrds4}
( \log A )'  = \frac{2}{\Delta'} \left(  1 - \frac{a^2}{L^2} \right)
\end{equation}
We compute 
\begin{equation}
\label{lambkerrds4}
\lambda (r,z) = \frac{\sqrt{\Delta_\theta} | \frac{\partial r}{\partial \theta} |}{\sqrt{\Delta} | \frac{\partial z}{\partial \rho} |}  =\frac{\tilde{B}'(z)}{\Xi A'(\rho)} \left( 1 + \frac{a^2}{L^2} \cos 2\theta \right)^{\frac{3}{2}}
\end{equation}
Writing the expansion of $\Delta(\rho)$ around $\rho=\rho_s$ as $\Delta(\rho) = \beta^2 - \alpha (\rho-\rho_s)^2 + \CO( (\rho-\rho_s)^3 )$ we can use the same analysis for the $A(\rho)$ function and the smoothness of $z(\rho,\theta)$ when $r <\beta$ as for static spherically symmetric dS$_4$ black holes. In particular this means the $z$ coordinate is a periodic coordinate and we choose the period to be $2\pi L$. One can now read off the domain structure. We see that $r=0$ for $\theta=0,\pi$ and $\rho=\rho_0,\rho_c$. Therefore, the domain structure of the Kerr-Newman-dS$_4$ black hole is given by four domains 
\begin{equation}
\label{domkerrds4}
\begin{array}{c}
W_1 = V_{(1)} \spa D_1 = [ - z_c , - z_0 ] \\[3mm]
W_2 = V_{(0)} + \Omega V_{(1)} \spa D_2 = [ - z_0 , z_0 ] \\[3mm]
W_3 = V_{(1)} \spa D_3 = [ z_0,  z_c ] \\[3mm]
W_4 = V_{(0)} + \Omega' V_{(1)}  \spa D_4 = [ z_c,  2\pi L - z_c ] 
\end{array}
\end{equation}
with $z_0$ and $z_c$ defined by requiring $0 < z_0/L < \pi/2$, $\pi/2 < z_c/L < \pi $, $B(z_0) = A(\rho_0)$ and $B(z_c) = A(\rho_c)$. Moreover, the angular velocities of the event horizon and cosmological horizon are given as
\begin{equation}
\Omega = \frac{a ( L^2 - \rho_0^2 )}{L^2 (\rho_0^2 + a^2 )} \spa \Omega' = \frac{a ( L^2 - \rho_c^2 )}{L^2 (\rho_c^2 + a^2 )}
\end{equation}
The domain structure \eqref{domkerrds4} is illustrated by Figure \ref{fig:domdsbh4}. For the domain lengths, as defined by \eqref{domlength}, we find using \eqref{kerrds4rz}, \eqref{aeqkerrds4} and \eqref{lambkerrds4}
\begin{equation}
\label{kerrdS4_D}
\begin{array}{c} \ds
|D_1| = |D_3| = \rho_c - \rho_0 \spa |D_2| = \frac{\Delta' (\rho_0)}{\Xi} = \frac{2}{\Xi} \left( \mu - \frac{\rho_0^3}{L^2} - \frac{q^2+a^2}{\rho_0} \right)  \\[5mm] \ds
|D_4| = - \frac{\Delta' (\rho_c)}{\Xi}  = \frac{2}{\Xi} \left( \frac{\rho_c^3}{L^2} + \frac{q^2+a^2}{\rho_c} - \mu \right)
\end{array}
\end{equation}

\subsubsection*{Multiple event horizons in dS$_4$}

\begin{figure}[h!]
\centerline{\includegraphics[scale=0.5]{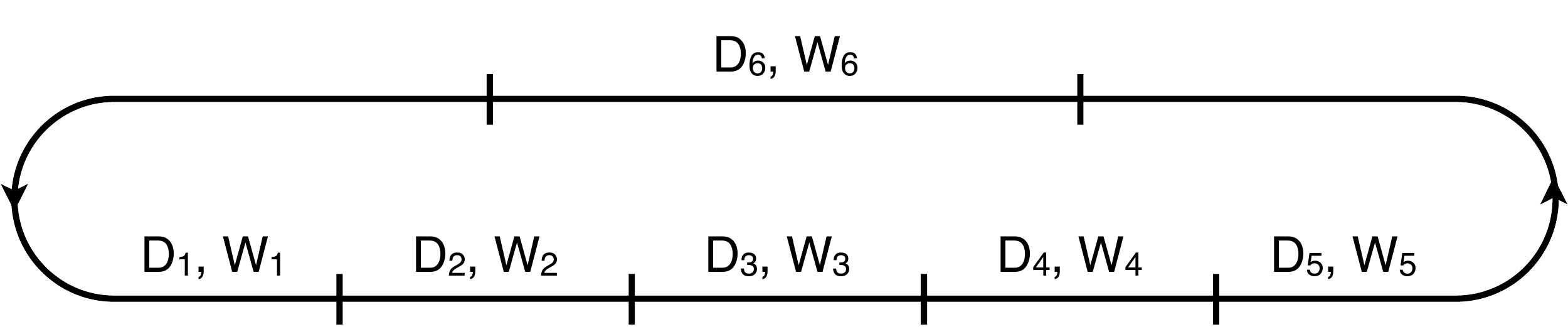}}
\vskip 0.2cm
\caption{\small Domain structure of two black holes in dS$_4$.}
\label{fig:dom2bhds}
\begin{picture}(0,0)(0,0)
\put(80,42){\footnotesize $\phi$-axis}
\put(125,42){\footnotesize Event horizon}
\put(208,42){\footnotesize $\phi$-axis}
\put(260,42){\footnotesize Event horizon}
\put(338,42){\footnotesize $\phi$-axis}
\put(180,88){\footnotesize Cosmological horizon}
\put(405,77){z}
\put(35,77){z}
\end{picture}
\end{figure}
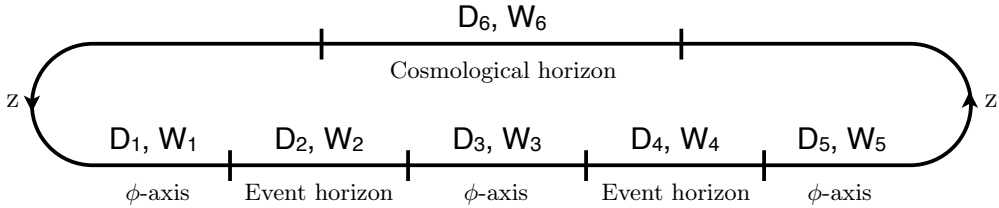

From the uniqueness theorems \cite{Israel:1967wq,Hawking:1972vc} it is clear that the only type of four-dimensional asymptotically flat black hole space-time one can have is one with a single connected event horizon of spherical topology. It seems likely that this is the case for asymptotically AdS$_4$ black hole space-times as well \cite{Anderson:2002xb}. However, for dS$_4$, using the coordinates of Eq.~\eqref{ds4}, two particles placed at $(\rho,\theta)=(\rho^*,0)$ and $(\rho,\theta)=(\rho^*,\pi)$ with $0 < \rho^* < L$ would accelerate away from each other, $i.e.$ we have a repulsive force in de Sitter space due to the positive cosmological constant. Therefore, adjusting the mass of the particles one could have equilibrium between the gravitational attraction and the repulsive force of the cosmological constant. Note that this configuration would still have rotational invariance in the $\phi$ direction since we placed the particles at the $\phi$-axis. We can now imagine putting two black holes in place of the two particles. Thus, from this argument it seems evident that there should exist regular asymptotically dS$_4$ black hole space-times with two disconnected event horizon of spherical topology in addition to the cosmological horizon. We have illustrated the domain structure of such a black hole space-time in Figure \ref{fig:dom2bhds}. Here $W_1 = W_3 = W_5 = \partial / \partial \phi$ and $W_2 = W_4 = W_6 = \partial / \partial t$ for a static configuration, however, such black hole space-times could presumably be generalized to 
stationary configurations as well.

\subsubsection*{Restrictions on horizon topologies in dS$_4$}

We consider now what restrictions on the horizon topology our domain structure analysis imposes on asymptotically dS$_4$ stationary black hole space-times.  Assuming the domain structure space $B$ to be a connected set, in this case meaning $B=S^1$, a domain with time-like direction, thus corresponding to a Killing horizon, will have a domain with direction $V_{(1)}$ on each side. From this we directly read off the topology of such an event horizon to be $S^2$. This argument applies to both the cosmological horizon as well as to any event horizon in the space-time.

\subsubsection*{Static spherically symmetric black holes in dS$_5$}

We consider here a rather general class of static spherically symmetric asymptotically dS$_5$ black hole space-times with metric of the form \eqref{genclass5} along with the following restrictions on the function $f(\rho)$ 
\begin{equation}
\label{staticBHds5}
\begin{array}{c} \ds
 \exists \rho_0, \rho_s, \rho_c:  \rho_c> \rho_s > \rho_0  > 0 \ , \ f(\rho_0)=f(\rho_c) =0 \ , \  ( \rho^4 f)' |_{\rho=\rho_s} = 0 \ ,   \\[3mm]
 (\rho^4 f)' > 0 \ \mbox{for} \ \rho_0 \leq \rho < \rho_s \ , \  (\rho^4 f)' < 0 \ \mbox{for} \ \rho_s < \rho \leq \rho_c \ ,
 \\[3mm] \ds  f (\rho) \simeq -\frac{\rho^2}{L^2} \ \ \mbox{for}\ \  \rho \gg \rho_c  
\end{array}
\end{equation}
It follows from this that we have a static and spherically symmetric black hole space-time with an event horizon at $\rho=\rho_0$ of $S^3$ topology and with a cosmological horizon at $\rho=\rho_c$ also with $S^3$ topology. The region $\rho_0 \leq \rho \leq \rho_c$ is the causal diamond region for which we would like to find the domain structure. Instead in the region $\rho \geq \rho_c$ we find the future null infinity $\CI^+$ for $\rho/L \rightarrow \infty$. 
Since the metric asymptote to \eqref{ds5} for $\rho/L \rightarrow \infty$ we should choose the three commuting Killing vectors as $V_{(0)} = \partial / \partial t$, $V_{(1)} = \partial / \partial \phi_1$ and $V_{(2)} = \partial / \partial \phi_2$ as in \eqref{ds5killing}. This ensures that the domain length \eqref{domlength} is well-defined. 

We use the ansatz \eqref{genans5} for canonical $(r,z)$ coordinates for the metric \eqref{genclass5} with \eqref{staticBHds5}. The metric is in the canonical form \eqref{genmetric} provided $A(\rho)$ solves \eqref{aeqads5}. Note that the requirements \eqref{staticBHds5} ensures that the function $\rho^4 f(\rho)$ only has a single extremum for $\rho_0 < \rho < \rho_c$ which is the maximum at $\rho=\rho_s$. We write the expansion around $\rho=\rho_s$ as $\rho^4 f(\rho) = 4 \hat{\beta}^2 - 4 \rho_s^2 \hat{\alpha}  (\rho-\rho_s)^2 + \CO( (\rho-\rho_s)^3)$ with $\hat{\alpha}, \hat{\beta} > 0$. With this one finds 
\begin{equation}
\label{thea_dsbh5}
A(\rho) \simeq \hat{c}  \left| 1 - \frac{\rho}{\rho_s}  \right|^{-\frac{1}{\hat{\alpha}}} \ \ \mbox{for}\ \ \rho \simeq \rho_s
\end{equation}
Thus, just as for pure dS$_5$ we have two patches for the $(r,z)$ coordinate system for the causal diamond defined by $\rho_0 \leq \rho \leq \rho_s$ and $\rho_s \leq \rho \leq \rho_c$. We also note that $0 \leq r \leq \hat{\beta}$ with the maximal value of $r$ reached in the point $(\rho,\theta)=(\rho_s,\pi/2)$. The topology of the $z$ coordinate is that of a periodic coordinate. We choose the period as $2\pi L^2$ and we make the following choice of $B(z)$ consistent with \eqref{thea_dsbh5} 
\begin{equation}
\label{theb_dsbh5}
\sin (\frac{z}{L^2}) \left| \cos ( \frac{z}{L^2} ) \right|^{-\frac{1}{\hat{\alpha}}} = A(\rho) \cos 2 \theta
\end{equation}
Just as for pure dS$_5$, this ensures a smooth function $z(\rho,\theta)$ for $z \neq \pm \pi L^2 /2$ and $r < \beta$. For $z \simeq \pm \pi L^2 /2$ we find $\cos (z/L^2) /  |\sin (z/L^2) |^{\hat{\alpha}} \simeq \hat{c}^{-\hat{\alpha}} ( 1 - \rho/ \rho_s) $ which ensures smoothness around $z = \pm \pi L^2 /2$.

\begin{figure}[h!]
\centerline{\includegraphics[scale=0.5]{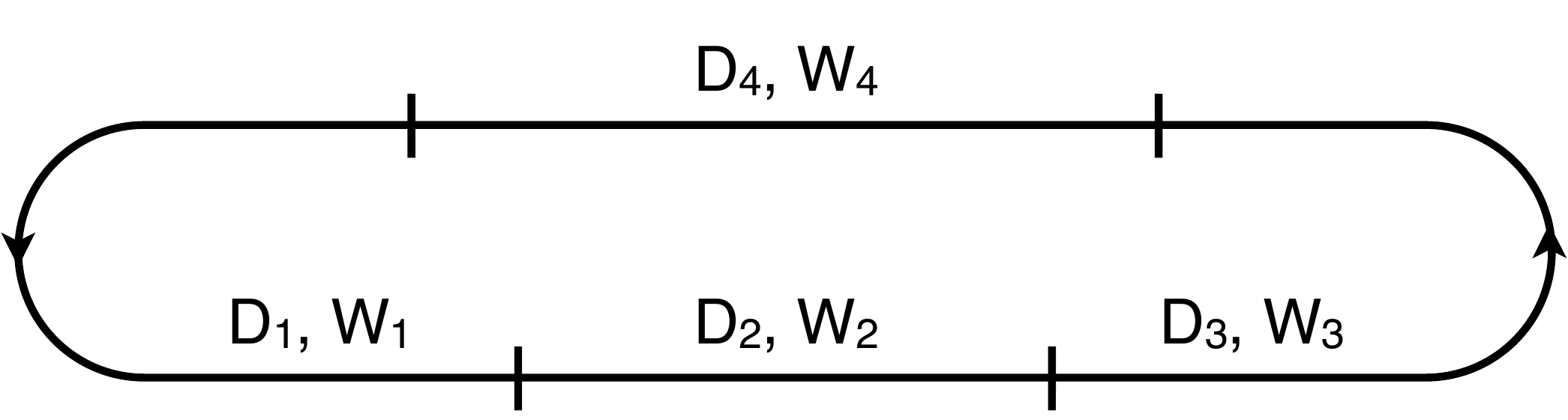}}
\vskip 0.2cm
\caption{\small Domain structure of black hole with $S^3$ topology in dS$_5$.}
\label{fig:domdsbh5}
\begin{picture}(0,0)(0,0)
\put(90,42){\footnotesize Fixed plane of $\phi_2$}
\put(195,42){\footnotesize Event horizon}
\put(285,42){\footnotesize Fixed plane of $\phi_1$}
\put(180,91){\footnotesize Cosmological horizon}
\put(376,78){z}
\put(64,78){z}
\end{picture}
\end{figure}

We can now read off the domain structure for the general class of static spherically symmetric asymptotically dS$_5$ black hole space-times with metrics \eqref{genclass5} and \eqref{staticBHds5}. Using \eqref{genans5}  we see that $r=0$ when $\theta=0,\pi /2$ and $\rho=\rho_0,\rho_c$. We have therefore the four domains
\begin{equation}
\label{domdsbh5}
\begin{array}{c}
W_1 = V_{(2)} \spa D_1 = [ - z_c , - z_0 ] \\[3mm]
W_2 = V_{(0)} \spa D_2 = [ - z_0 , z_0 ] \\[3mm]
W_3 = V_{(1)} \spa D_3 = [ z_0,  z_c ] \\[3mm]
W_4 = V_{(0)} \spa D_4 = [ z_c,  2\pi L^2 - z_c ] 
\end{array}
\end{equation}
where we defined $z_0$ and $z_c$ by $0 < z_0/L^2 < \pi/2$, $\pi/2 < z_c/L^2 < \pi $, $B(z_0) = A(\rho_0)$ and $B(z_c) = A(\rho_c)$.
The domains $D_1$ and $D_3$ are the fixed planes for the $\phi_2$ and $\phi_1$ rotation angles, respectively, while $D_2$ and $D_4$ are the black hole event horizon and the cosmological horizon, respectively. The domain structure \eqref{domdsbh5} is illustrated in Figure \ref{fig:domdsbh5}.
Using \eqref{genans5}, \eqref{aeqads5} and \eqref{lamb5} we easily calculate the invariant domain lengths defined by Eq.~\eqref{domlength}
\begin{equation}
|D_1| = |D_3| = \frac{1}{2} ( \rho_c^2 - \rho_0^2 ) \spa |D_2| = \frac{\rho_0^3}{4} f'(\rho_0) \spa |D_4| = - \frac{\rho_c^3}{4} f'(\rho_c)
\end{equation}
The total invariant length is thus $\rho_c^2 - \rho_0^2 + \frac{1}{4} \rho_0^3 f'(\rho_0) - \frac{1}{4} \rho_c^3 f'(\rho_c)$.

We note again that we found the domain structure above without the need to know what types of matter that should be present.  A particularly example is the Reissner-Nordstr\" om-dS$_5$ black hole which is a static electrically charged spherically symmetric black hole that is a solution to 5D Einstein-Maxwell gravity with a cosmological constant $\Lambda = 6/ L^2$. The metric in the solution is given by \eqref{genclass5} with
\begin{equation}
\label{RNdS5}
f(\rho) = 1 - \frac{\rho^2}{L^2} -\frac{2\mu}{\rho^2} + \frac{q^2}{\rho^4}
\end{equation}
where $\mu$ is proportional to the mass and $q$ to the electric charge of the black hole. Consider now a choice of parameters $L$, $\mu$ and $q$ for which the requirements \eqref{staticBHds5} are met, with $\rho_0$ and $\rho_c$ marking the event horizon and the cosmological horizon, respectively.
From the above we get then that
the domain structure is given by \eqref{domdsbh5} with 
\begin{equation}
|D_1| = |D_3| = \frac{1}{2} ( \rho_c^2 - \rho_0^2 ) \spa |D_2| = \mu - \frac{\rho_0^4}{2L^2} - \frac{q^2}{\rho_0^2} \spa |D_4| =  \frac{\rho_c^4}{2L^2} - \mu + \frac{q^2}{\rho_c^2}
\end{equation}

\subsubsection*{Kerr-dS$_5$ black hole}

The Kerr-dS$_5$ black hole is a stationary asymptotically dS$_5$ black hole with angular momenta and with an event horizon homeomorphic to a three-sphere \cite{Hawking:1998kw,Gibbons:2004uw}. It is a solution to the Einstein equations with a cosmological constant $\Lambda = 6 / L^2$. 
The metric can be written as
\begin{equation}
\label{kerrds5}
\begin{array}{rcl}
\ds ds^2 &=& \ds - \frac{\Delta}{\Sigma} \Big( dt - \sum_{i=1}^2 \frac{a_i \mu_i^2}{\Xi_i} d\phi_i \Big)^2 + \frac{\Delta_\theta}{\Sigma} \sum_{i=1}^2 \mu_i^2 \Big( a_i dt - \frac{\rho^2 + a_i^2}{\Xi_i} d\phi_i \Big)^2 \\[2mm] \ds && \ds + \frac{ L^2-\rho^2 }{L^2 \rho^2 \Sigma} \Big( a_1 a_2 dt - \sum_{i=1}^2 \frac{a_i (\rho^2 + a_i^2 )}{\Xi_i} \mu_i^2 d\phi_i \Big)^2 + \Sigma \left[ \frac{d\rho^2}{\Delta} + \frac{d\theta^2}{\Delta_\theta} \right]
\end{array}
\end{equation}
with the definitions
\begin{equation}
\begin{array}{c} \ds
\Delta = \frac{1}{\rho^2} \Big(1 - \frac{\rho^2}{L^2} \Big) \prod_{i=1}^2 ( \rho^2 + a_i^2 ) - 2 \mu  \spa \Sigma = \rho^2 + \sum_{i=1}^2 a_i^2 (1- \mu_i^2)  \\[2mm]  \ds \Delta_\theta = 1 + \sum_{i=1}^2 \frac{a_i^2}{L^2} (1-\mu_i^2 ) \spa
\ds  \Xi_i = 1 + \frac{a_i^2}{L^2} \spa \mu_1 = \sin \theta \spa \mu_2 = \cos \theta 
\end{array}
\end{equation}
The parameters $a_1$ and $a_2$ are the rotation parameters. For $a_1=a_2=0$ we regain the Schwarzschild-dS$_5$ black hole given by Eqs.~\eqref{genclass5} and \eqref{RNdS5} with $q=0$. 
The metric \eqref{kerrds5} can be obtained from the metric \eqref{kerrads5} of the Kerr-AdS$_5$ black hole by analytic continuation $L^2 \rightarrow - L^2$.

We assume the parameters $L$, $\mu$, $q$ and $a$ are such that 
\begin{equation}
\begin{array}{c} \ds
 \exists \rho_0, \rho_s, \rho_c:  \rho_c> \rho_s > \rho_0  > 0 \ , \ \Delta(\rho_0)=\Delta(\rho_c) =0 \ , \  ( \rho^2 \Delta)' |_{\rho=\rho_s} = 0 \ ,   \\[3mm]
 (\rho^2 \Delta)' > 0 \ \mbox{for} \ \rho_0 \leq \rho < \rho_s \ , \  (\rho^2 \Delta)' < 0 \ \mbox{for} \ \rho_s < \rho \leq \rho_c 
 \end{array}
\end{equation}
At $\rho=\rho_0$ we have an event horizon with $S^3$ topology and at $\rho=\rho_c$ a cosmological horizon also with $S^3$ topology. The causal diamond within which we consider the domain structure is defined by $\rho_0 \leq \rho \leq \rho_c$. The metric \eqref{kerrds4} can be analytically extended for $\rho \geq \rho_c$ and we find the asymptotic region for $\rho \gg \rho_c$ near future null infinity $\CI^+$. 

Finding the domain structure basically consists in combining the analysis of static spherically symmetric dS$_5$ black holes \eqref{staticBHds5} with the analytical continuation $L^2 \rightarrow - L^2$ of the analysis of the Kerr-AdS$_5$ black hole \eqref{kerrads5}. We will therefore be brief in the following. 

The correct choice of the Killing vectors are
\begin{equation}
\label{kerrds5kill}
V_{(0)} = \frac{\partial}{\partial t} + \sum_{i=1}^2 \frac{a_i}{L^2} \frac{\partial}{\partial \phi_i} \spa V_{(1)} = \frac{\partial}{\partial \phi_1} \spa V_{(2)} = \frac{\partial}{\partial \phi_2} 
\end{equation}
since this ensures that $V_{(0)}$ asymptotes to the Killing vector field $\partial / \partial t$ for the metric \eqref{ds5} of dS$_5$ for $\rho\rightarrow \infty$, $i.e.$ near null infinity $\CI^+$. The coordinates $(r,z)$ give a metric on the canonical form \eqref{genmetric} with
\begin{equation}
\label{kerrds5rz}
r = \frac{\rho \sqrt{\Delta}}{2\Xi_1 \Xi_2} \sqrt{\Delta_\theta} \sin 2\theta \spa \tilde{B} (z) = A(\rho) g(\theta)
\end{equation}
provided we have
\begin{equation}
\label{aeqkerrds5}
(\log A)' =  \frac{8 \rho^2 \sqrt{\epsilon^2 + \Xi_1 \Xi_2} }{( \rho^2 \Delta )'}
\spa
g(\theta) = - \frac{4}{3\epsilon}
\frac{3 \epsilon \cos 2\theta - \Xi_1 - \Xi_2 + 2\sqrt{ \epsilon^2 + \Xi_1 \Xi_2}  }{3 \epsilon \cos 2\theta - \Xi_1 - \Xi_2 - 2\sqrt{ \epsilon^2 +  \Xi_1 \Xi_2} } 
\end{equation}
with $\epsilon L^2 = a_2^2 - a_1^2$. Note that $g'(\theta) < 0$ for $0 < \theta < \pi/2$, $g'(0)=g'(\pi/2) = 0$ and that for $\epsilon \rightarrow 0$ we get $g(\theta) =  \cos (2 \theta ) / \Xi_1$.
We compute 
\begin{equation}
\label{lambkerrds5}
\lambda (r,z) = \frac{\sqrt{\Delta_\theta} | \frac{\partial r}{\partial \theta} |}{\sqrt{\Delta} | \frac{\partial z}{\partial \rho} |}  = \frac{\rho \tilde{B}' (z) }{ A'(\rho) } \frac{\sqrt{\Delta_\theta} ( \sqrt{\Delta_\theta} \sin 2\theta )'}{2\Xi_1 \Xi_2 g(\theta)}
\end{equation}
Writing the expansion of $\rho^2 \Delta(\rho)$ around $\rho=\rho_s$ as $\rho^2 \Delta(\rho) = 4 \hat{\beta}^2 - 4 \rho_s^2 \hat{\alpha}  (\rho-\rho_s)^2 + \CO( (\rho-\rho_s)^3)$ we can use the same analysis for the $A(\rho)$ function and the smoothness of $z(\rho,\theta)$ when $r <\hat{\beta}$ as for static spherically symmetric dS$_5$ black holes. In particular this means the $z$ coordinate is a periodic coordinate and we choose the period to be $2\pi L^2$. We can readily read off the domain structure. We see that $r=0$ for $\theta=0,\pi /2$ and $\rho=\rho_0,\rho_c$. Therefore, the domain structure of the Kerr-dS$_5$ black hole is given by four domains 
\begin{equation}
\label{domkerrds5}
\begin{array}{c}
W_1 = V_{(2)} \spa D_1 = [ - z_c , - z_0 ] \\[3mm]
W_2 = V_{(0)} + \Omega_1 V_{(1)} + \Omega_2 V_{(2)} \spa D_2 = [ - z_0 , z_0 ] \\[3mm]
W_3 = V_{(1)} \spa D_3 = [ z_0,  z_c ] \\[3mm]
W_4 = V_{(0)} + \Omega'_1 V_{(1)}  +  \Omega'_2 V_{(2)}  \spa D_4 = [ z_c,  2\pi L - z_c ] 
\end{array}
\end{equation}
with $z_0$ and $z_c$ defined by requiring $0 < z_0/L < \pi/2$, $\pi/2 < z_c/L < \pi $, $B(z_0) = A(\rho_0)$ and $B(z_c) = A(\rho_c)$. Moreover, the angular velocities of the event horizon and cosmological horizon are given as
\begin{equation}
\Omega_i = \frac{a_i ( L^2 - \rho_0^2 )}{L^2 (\rho_0^2 + a_i^2 )} \spa \Omega'_i = \frac{a_i ( L^2 - \rho_c^2 )}{L^2 (\rho_c^2 + a_i^2 )}
\end{equation}
The domain structure \eqref{domkerrds5} is illustrated by Figure \ref{fig:domdsbh5}. For the domain lengths, as defined by \eqref{domlength}, we find using \eqref{kerrds5rz}, \eqref{aeqkerrds5} and \eqref{lambkerrds5}
\begin{equation}
\label{kerrdS5_D}
\begin{array}{c} \ds
|D_1| = \frac{\rho_c^2 - \rho_0^2}{2\Xi_1}  \spa   |D_2| =  \frac{\rho_0 \Delta' (\rho_0) }{4 \Xi_1 \Xi_2 } = \frac{1}{\Xi_1 \Xi_2} \left[ \mu - \frac{\rho_0^4}{2L^2} - \frac{a_1^2+a_2^2}{2} - \frac{a_1^2a_2^2}{\rho_0^2} + \frac{a_1^2a_2^2}{2L^2}    \right]  \\[5mm] \ds
|D_3| = \frac{\rho_c^2 - \rho_0^2 }{2 \Xi_2}  \spa |D_4| = - \frac{\rho_c \Delta' (\rho_c) }{4 \Xi_1 \Xi_2 } = \frac{1}{\Xi_1 \Xi_2} \left[  \frac{\rho_c^4}{2L^2} + \frac{a_1^2+a_2^2}{2} + \frac{a_1^2a_2^2}{\rho_c^2} - \frac{a_1^2a_2^2}{2L^2} - \mu   \right]
\end{array}
\end{equation}

\subsubsection*{Black rings and multiple event horizons in dS$_5$}

Following the discussion of the domain structure of asymptotically AdS$_5$ black hole space-times we should expect that all the types of five-dimensional asymptotically flat black hole space-times (with two rotational Killing vector fields) can be found for asymptotically dS$_5$ black hole space-times as well, with the addition of the cosmological horizon. $E.g.$ the black ring, the black Saturn, etc., should all be present with a positive cosmological constant as well. To find the domain structure of these one should merely take the domain structure of the asymptotically flat black hole space-time and connect the $z \rightarrow -\infty$ and $z \rightarrow \infty$ ends adding also the domain corresponding to the cosmological horizon. 

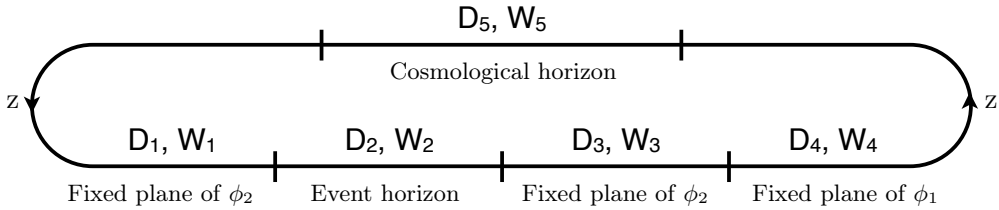
\begin{figure}[h!]
\centerline{\includegraphics[scale=0.5]{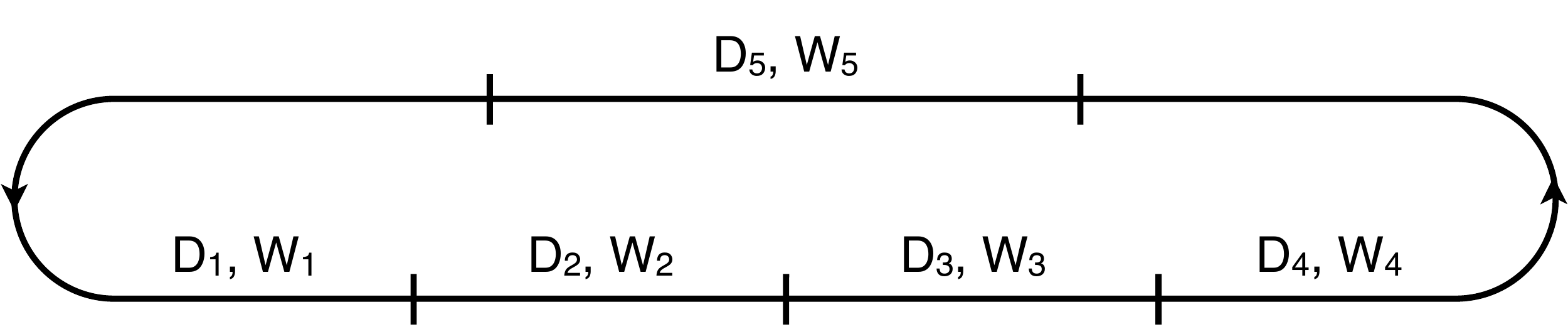}}
\vskip 0.2cm
\caption{\small Domain structure of black ring in dS$_5$.}
\label{fig:domdsbr}
\begin{picture}(0,0)(0,0)
\put(58,42){\footnotesize Fixed plane of $\phi_2$}
\put(150,42){\footnotesize Event horizon}
\put(230,42){\footnotesize Fixed plane of $\phi_2$}
\put(317,42){\footnotesize Fixed plane of $\phi_1$}
\put(180,88){\footnotesize Cosmological horizon}
\put(405,77){z}
\put(35,77){z}
\end{picture}
\end{figure}

A particular example is that of the black ring in dS$_5$. The domain structure for such a black hole space-time is illustrated in Figure \ref{fig:domdsbr} where $W_1 =W_3 = \partial / \partial \phi_2$ and $W_4 = \partial / \partial \phi_1$. Evidence for such black hole space-times have been found using the blackfold method \cite{Caldarelli:2008pz}. In particular, a configuration was found where the space-time is static, $i.e.$ $W_2 = W_5 = \partial / \partial t$. Note that physically this makes sense since the positive cosmological constant gives a repulsive force, thus it is the same type of equilibrium as that of two black holes in dS$_4$.

\subsubsection*{Restrictions on horizon topologies in dS$_5$}

We can now consider what restrictions on the horizon topology our domain structure analysis imposes on asymptotically dS$_5$ stationary black hole space-times with two rotational Killing vector fields. Consider again a domain with time-like direction, thus corresponding to a Killing horizon, and assume that the domain structure space $B$ is a connected set. The most general situation for the domain is that one has a domain $D'$ with space-like direction $V'$ on its left side and a domain $D''$ with space-like direction $V''$ on its right side.  It follows from Theorem \ref{theo:dom} of Section \ref{sec:generaldom} that $V'$ and $V''$ should be of the form $V' = p' V_{(1)} + q' V_{(2)}$ and $V'' = p'' V_{(1)} + q'' V_{(2)}$ where $p'$ and $q'$ are mutually prime numbers, and the same for $p''$ and $q''$. It is straightforward to infer from this that the horizon topology is restricted to be either $S^3$, $S^2 \times S^1$ or a Lens-space. This argument applies to both the cosmological horizon as well as to any event horizon in the space-time.

\section{Domain structure of higher-dimensional space-times}
\label{sec:higherdim}

In this section we briefly consider the domain structure of asymptotically AdS and dS black hole space-times in higher than five space-time dimensions. For a $D$-dimensional space-time with $D \geq 6$ one can at most have $D-3$ commuting Killing vector fields if the space-time is asymptotically flat. The same holds for asymptotically AdS and dS space-times. The domain structure for asymptotically flat black hole space-times with $D \geq 6$ was developed in \cite{Harmark:2009dh}. Rather than living on $\R$, the domain structure lives on $\R^m$. Specifically, for $D=6$ and $D=7$, with the maximal number of commuting rotational Killing vector fields present, the domain structure lives on $\R^2$. %
\footnote{Note that the generalization of the domain structure to a higher-dimensional space $\R^m$ gives a certain hierarchy of submanifolds that generalize the intervals and interval end points for space-times with $D-2$ commuting Killing vector fields. See \cite{Harmark:2009dh} for a detailed exposition of this.}

To illustrate the generalization of the domain structure to higher dimensional space-times with a cosmological constant we focus in the following on the domain structure of six and seven dimensional black hole space-times with the maximal number of commuting rotational Killing vector fields. The canonical metric is of the form \cite{Harmark:2009dh}%
\footnote{For sake of simplicity we do not write the most general form of the metric with mixed terms between the Killing directions and $(r,z^1,z^2)$ directions since this is not needed for the examples below.}
\begin{equation}
\label{canform67}
ds^2 = \sum_{i,j=1}^{D-3} G_{ij} dx^i dx^j + e^{2\nu} \Big( dr^2 + \sum_{a,b=1}^2 \lambda_{ab} dz^a dz^b \Big) \spa r = \sqrt{ | \det G_{ij} | }
\end{equation}
where the Killing vector fields of the metric $V_{(1)},...,V_{(D-3)}$ are of the form \eqref{killv}.
The domain structure then lives at the $r=0$ submanifold of the three-dimensional orbit space which can be parameterized by $(r,z^1,z^2)$ with the above metric restricted to $dx^i=0$. Thus, the domain structure lives on a two-dimensional space $B$ parameterized by $(z^1,z^2)$ and with metric $ds^2 = \sum_{a,b=1}^2 \lambda_{ab} |_{r=0} dz^a dz^b$. For a given black hole space-time one now has two types of invariants \cite{Harmark:2009dh}: 1) The topological division of the two-dimensional domain structure space $B$ into domains $B = D_1 \cup \cdots \cup D_N$, and 2) The area of the domains $D_i$, as measured with
\begin{equation}
\label{domarea}
|D_i| = \int_{D_i} \lambda \, dz^1 dz^2   \spa \lambda \equiv \sqrt{\det \lambda_{ab}}
\end{equation}
One can easily work out that the domain structure of asymptotically AdS$_D$ black hole space-times with $D=6,7$ topologically is the same as the domain structure of asymptotically flat black hole space-times with the same topological structure of event horizons present in the space-time. We therefore restrict ourselves to the cases of asymptotically dS$_6$ and dS$_7$ space-times below.

To consider asymptotically dS$_6$ space-times we start with the general six-dimensional spherically symmetric metric
\begin{equation}
\label{6dmet}
ds^2 = - f dt^2 + \frac{d\rho^2}{f} + \rho^2 ( \sin^2 \theta d\phi_1^2 + \cos^2 \theta \sin^2 \psi d\phi_2^2 + d\theta^2 + \cos^2 \theta d\psi^2 )
\end{equation}
where $f=f(\rho)$, $0 \leq \theta \leq \pi /2$ and $0 \leq \psi \leq \pi$. We make the following choice of commuting Killing vector fields
\begin{equation}
\label{6dkill}
V_{(0)} = \frac{\partial}{\partial t} \spa V_{(1)} = \frac{\partial}{\partial \phi_1}  \spa V_{(2)} = \frac{\partial}{\partial \phi_2} 
\end{equation}
For $f = 1 - \rho^2 / L^2$ and $\rho < L$ this metric describes the (Southern) causal diamond region of dS$_6$ while for $\rho >L$ it describes the region with $\CI^+$. The metric \eqref{6dmet} is transformed to the canonical form \eqref{canform67} provided the $(r,z^1,z^2)$ coordinates are given by 
\begin{equation}
\label{6drz}
r = \frac{1}{2} \sqrt{ \rho^4 f} \sin 2\theta \sin \psi \spa B_1(z^1) = b_1 (\rho) \cos \theta \cos \psi \spa B_2 (z^2) = b_2 (\rho) \cos 2\theta
\end{equation}
with
\begin{equation}
\label{6dbeqs}
(\log b_2 )' = 2 (\log b_1)' = \frac{8 \rho^2}{(\rho^4 f)'}
\end{equation}
We compute
\begin{equation}
\label{6dlambda}
\lambda = \frac{4 \rho^4 B_1' B_2'}{(\rho^4 f)' b_1' b_2'}
\end{equation}
Consider now dS$_6$ corresponding to $f= 1- \rho^2 /L^2$ and a positive cosmological constant $\Lambda = 10/ L^2$.
Since $\rho^4 f(\rho)$ reaches a maximum at $\rho=\rho_s \equiv \sqrt{3} L / \sqrt{2} $ we have that $r$ reaches a maximal possible value at $(\rho,\theta,\psi)=(\rho_s,\pi /4, \pi /2)$. Studying the surfaces of constant $r$ in the $(\rho,\theta,\psi)$ space one finds that they have $S^2$ topology. We find $b_1(\rho) \propto \rho / \sqrt{| 1 - \rho^2 / \rho_s^2 |}$ and $b_2 (\rho)\propto \rho^2 / | 1 - \rho^2 / \rho_s^2 |$. Thus, it is consistent to take both $z_1$ and $z_2$ to be periodic coordinates. Specifically, we choose 
\begin{equation}
\frac{\sin z_1}{\sqrt{| \cos z_1 |}} = \frac{\rho}{\sqrt{| \rho_s^2 - \rho^2 |}} \cos \theta \cos \psi \spa \tan z_2 =  \frac{\rho^2}{\rho_s^2 - \rho^2} \cos 2\theta
\end{equation}
We require in addition that $\cos z_1$ and $\rho_s - \rho$ have the same sign, ensuring smoothness of $z_i(\rho,\theta,\psi)$ near $\rho=\rho_s$. However, the $z_1$ and $z_2$ coordinates cover twice a surface of constant $r$. To only cover the surface once we restrict ourselves to $- \pi/2 \leq z_1 \leq \pi /2$. This is in accordance with the surfaces of constant $r$ having $S^2$ topology. 
Moreover, the points $z_1 = \pm \pi/2$ behave like poles in that given $z_1 = \pm \pi/2$ a point $(\rho,\theta,\psi)$ only depends on the value of $r$ and not on the value of $z_2$. 

\begin{figure}[h!]
\centerline{\includegraphics[scale=0.6]{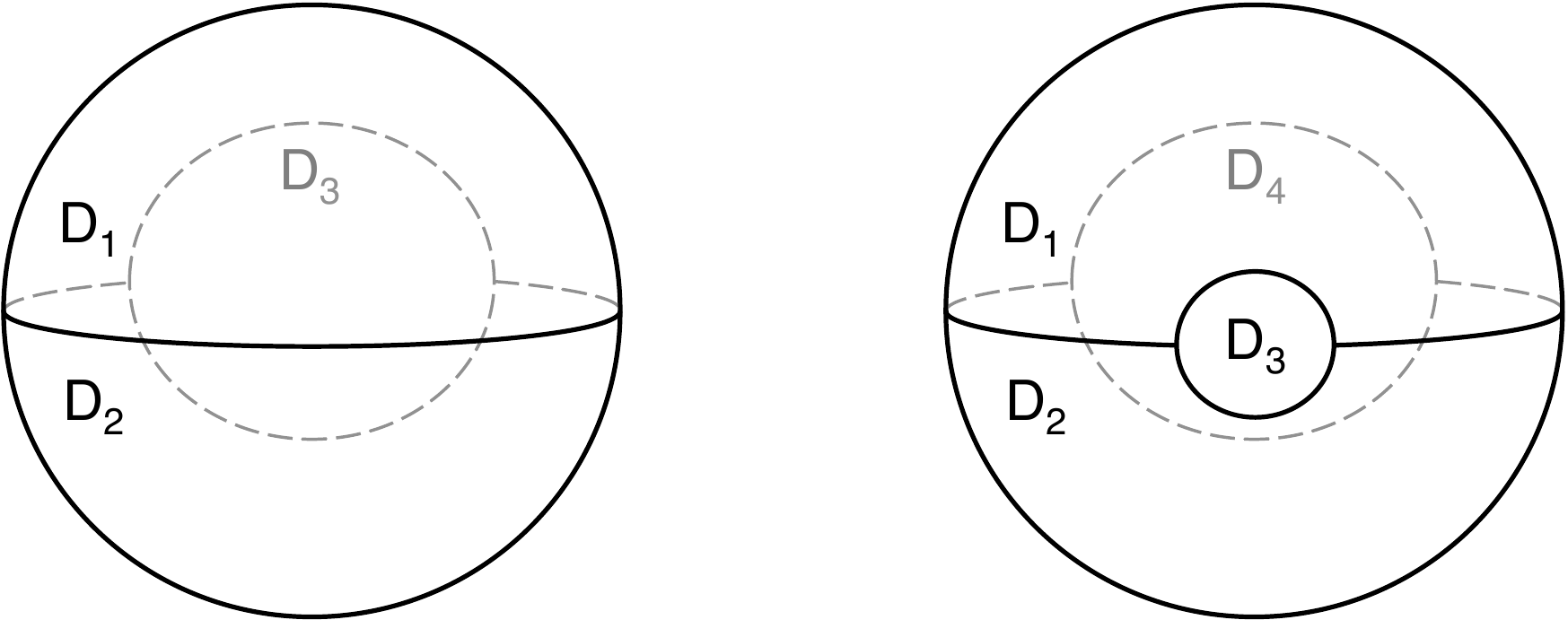}}
\caption{\small Left side: Domain structure of dS$_6$. Right side: Domain structure of black hole with $S^4$ topology in dS$_6$.}
\label{fig:ds6d}
\end{figure}

The domain structure of dS$_6$ (in the Southern causal diamond patch) is given by three domains, $D_1$ corresponding to $\theta=0$, $D_2$ to $\psi=0,\pi$ and $D_3$ to $\rho=L$, with the domain directions
\begin{equation}
W_1 = \frac{\partial}{\partial \phi_1} \spa W_2 = \frac{\partial}{\partial \phi_2} \spa W_3 =  \frac{\partial}{\partial t}
\end{equation}
The domains $D_1$ and $D_2$ corresponds to the hyperplanes of fixed points for the rotation angles $\phi_1$ and $\phi_2$, respectively, while $D_3$ corresponds to the cosmological horizon. The three domains constitutes a division of the two-sphere $S^2 = D_1 \cup D_2 \cup D_3$. We have illustrated the domain structure of dS$_6$ in Figure \ref{fig:ds6d}. Using Eqs.~\eqref{6drz}, \eqref{6dbeqs} and \eqref{6dlambda} it is straightforward to compute the domain areas
\begin{equation}
| D_1 | = | D_2 | = | D_3 | = \frac{2}{3} L^3
\end{equation}
as measured by \eqref{domarea}.

Turning to asymptotically dS$_6$ space-times we consider the following class of static spherically symmetric black hole space-times given by the metric \eqref{6dmet} with
\begin{equation}
\label{staticBHds6}
\begin{array}{c} \ds
 \exists \rho_0, \rho_s, \rho_c:  \rho_c> \rho_s > \rho_0  > 0 \ , \ f(\rho_0)=f(\rho_c) =0 \ , \  ( \rho^4 f)' |_{\rho=\rho_s} = 0 \ ,   \\[3mm]
 (\rho^4 f)' > 0 \ \mbox{for} \ \rho_0 \leq \rho < \rho_s \ , \  (\rho^4 f)' < 0 \ \mbox{for} \ \rho_s < \rho \leq \rho_c \ ,
 \\[3mm] \ds  f (\rho) \simeq -\frac{\rho^2}{L^2} \ \ \mbox{for}\ \  \rho \gg \rho_c  
\end{array}
\end{equation}
The choice of commuting Killing vector fields is \eqref{6dkill}. For this general class of space-times we find four domains, $D_1$ corresponding to $\theta=0$, $D_2$ to $\psi=0,\pi$, $D_3$ to $\rho=\rho_0$ and $D_4$ to $\rho= \rho_c$, with the domain directions
\begin{equation}
W_1 = \frac{\partial}{\partial \phi_1} \spa W_2 = \frac{\partial}{\partial \phi_2} \spa W_3 =  W_4 = \frac{\partial}{\partial t}
\end{equation}
We have illustrated the domain structure of this general class of asymptotically dS$_6$, static and spherically symmetric black hole space-times in Figure \ref{fig:ds6d}.
The domains $D_1$ and $D_2$ correspond to hyperplanes of fixed points for the rotation angles $\phi_1$ and $\phi_2$, respectively, while $D_3$ and $D_4$ correspond to the black hole event horizon and the cosmological horizon, respectively. Using Eqs.~\eqref{6drz}, \eqref{6dbeqs} and \eqref{6dlambda} we compute the domain areas
\begin{equation}
|D_1| = | D_2 | = \frac{2}{3} ( \rho_c^3 - \rho_0^3 ) \spa |D_3| = \frac{\rho_0^4}{3} f' (\rho_0) \spa |D_4| = - \frac{\rho_c^4}{3} f' (\rho_c) 
\end{equation}
as measured by \eqref{domarea}. Considering the specific case of a Reissner-Nordstr\" om-dS$_6$ black hole, which is a static electrically charged spherically symmetric black hole that is a solution to 6D Einstein-Maxwell gravity with a cosmological constant $\Lambda = 10/ L^2$, we have
\begin{equation}
\label{6df}
f(\rho) = 1 - \frac{\rho^2}{L^2} - \frac{2\mu}{\rho^3} + \frac{q^2}{\rho^6}
\end{equation}
and we find the domain areas
\begin{equation}
|D_1| = | D_2 | = \frac{2}{3} ( \rho_c^3 - \rho_0^3 ) \spa |D_3| = 2 \mu - \frac{2\rho_0^5}{3L^2} - \frac{2q^2}{\rho_0^3}  \spa |D_4| =  \frac{2\rho_c^5}{3L^2} - 2 \mu + \frac{2q^2}{\rho_c^3} 
\end{equation}

We now turn to asymptotically dS$_7$ space-times.  We start with the general seven-dimensional spherically symmetric metric
\begin{equation}
\label{7dmet}
ds^2 = - f dt^2 + \frac{d\rho^2}{f} + \rho^2 ( \sin^2 \theta d\phi_1^2 + \cos^2 \theta \sin^2 \psi d\phi_2^2  + \cos^2 \theta \cos^2 \psi d\phi_3^2 + d\theta^2 + \cos^2 \theta d\psi^2 )
\end{equation}
where $f=f(\rho)$, $0 \leq \theta \leq \pi /2$ and $0 \leq \psi \leq \pi /2$. We make the following choice of commuting Killing vector fields
\begin{equation}
\label{7dkill}
V_{(0)} = \frac{\partial}{\partial t} \spa V_{(1)} = \frac{\partial}{\partial \phi_1}  \spa V_{(2)} = \frac{\partial}{\partial \phi_2}   \spa V_{(3)} = \frac{\partial}{\partial \phi_3} 
\end{equation}
For $f = 1 - \rho^2 / L^2$ and $\rho < L$ this metric describes the (Southern) causal diamond region of dS$_7$ while for $\rho >L$ it describes the region with $\CI^+$. The metric \eqref{7dmet} is transformed to the canonical form \eqref{canform67} provided the $(r,z^1,z^2)$ coordinates are given by 
\begin{equation}
\label{7drz}
r = \frac{1}{2} \sqrt{ \rho^6 f} \sin \theta \cos^2 \theta \sin 2\psi \spa B_1(z^1) = b_1 (\rho) \cos^2 \theta \cos 2\psi \spa B_2 (z^2) = b_2 (\rho) ( 3 \cos^2\theta - 2)
\end{equation}
with
\begin{equation}
\label{7dbeqs}
( \log b_1 )' = (\log b_2 )' = \frac{12 \rho^4}{(\rho^6 f)'}
\end{equation}
We compute
\begin{equation}
\label{7dlambda}
\lambda = \frac{3 \rho^7 B_1' B_2'}{(\rho^6 f)' b_1' b_2'}
\end{equation}
Consider now dS$_7$ corresponding to $f= 1- \rho^2 /L^2$ and a positive cosmological constant $\Lambda = 15/ L^2$.
Since $\rho^6 f(\rho)$ reaches a maximum at $\rho=\rho_s \equiv \sqrt{3}  L /2  $ we have that $r$ reaches a maximal possible value at $(\rho,\theta,\psi)=(\rho_s, \arccos (\sqrt{2/3}), \pi /4)$. Studying the surfaces of constant $r$ in the $(\rho,\theta,\psi)$ space one finds that they have $S^2$ topology. We find $b_1(\rho) \propto \rho^2 / | 1 - \rho^2 / \rho_s^2 |$ and $b_2 (\rho)\propto \rho^2 / | 1 - \rho^2 / \rho_s^2 |$. Thus, it is consistent to take both $z_1$ and $z_2$ to be periodic coordinates. Specifically, we choose 
\begin{equation}
\tan z_1 = \frac{\rho^2}{\rho_s^2 - \rho^2} \cos^2 \theta \cos 2\psi \spa \tan z_2 =  \frac{\rho^2}{\rho_s^2 - \rho^2} (3 \cos^2 \theta -2)
\end{equation}
However, the $z_1$ and $z_2$ coordinates cover twice a surface of constant $r$. To only cover the surface once we restrict ourselves to $- \pi/2 \leq z_1 \leq \pi /2$. This is in accordance with the surfaces of constant $r$ having $S^2$ topology.  Moreover, the points $z_1 = \pm \pi/2$ behave like poles in that given $z_1 = \pm \pi/2$ a point $(\rho,\theta,\psi)$ only depends on the value of $r$ and not on the value of $z_2$. 

\begin{figure}[h!]
\centerline{\includegraphics[scale=0.6]{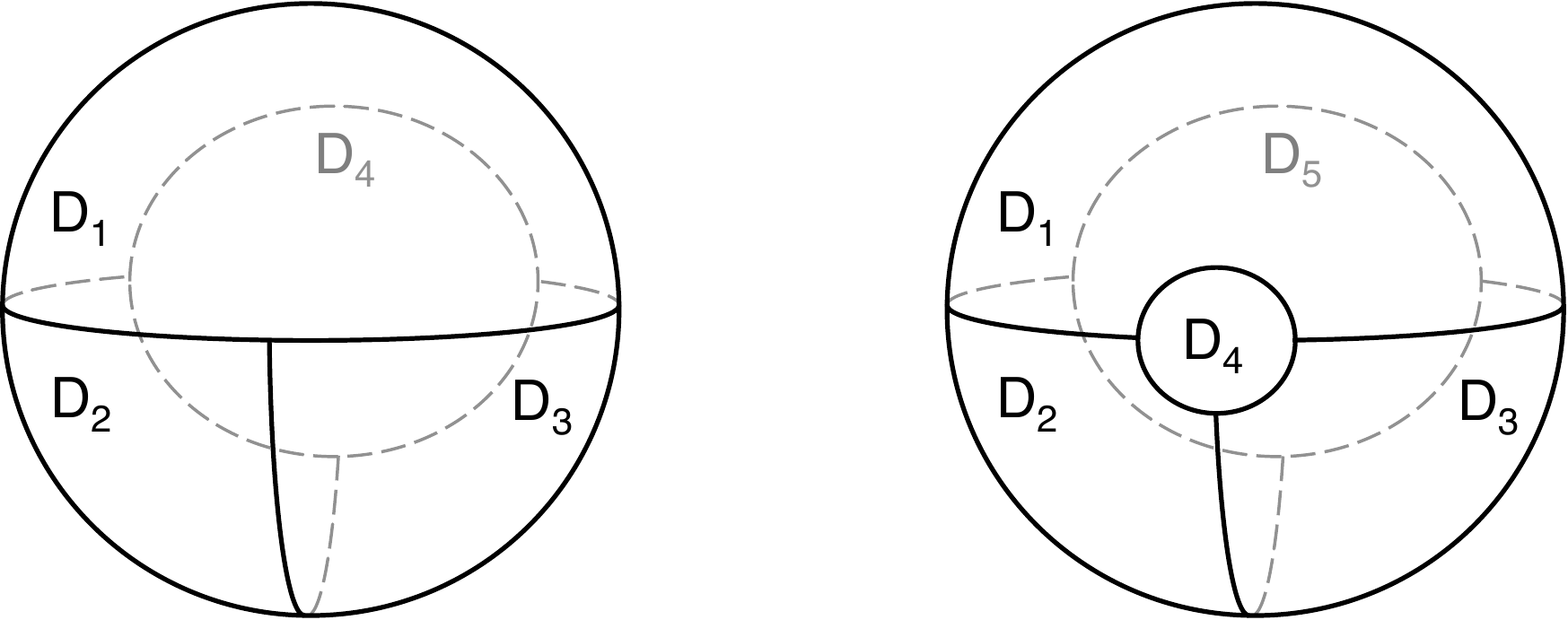}}
\caption{\small Left side: Domain structure of dS$_7$. Right side: Domain structure of black hole with $S^5$ topology in dS$_7$.}
\label{fig:ds7d}
\end{figure}

The domain structure of dS$_7$ (in the Southern causal diamond patch) is given by four domains, $D_1$ corresponding to $\theta=0$, $D_2$ to $\psi=0$, $D_3$ to $\psi=\pi/2$ and $D_4$ to $\rho=L$, with the domain directions
\begin{equation}
W_1 = \frac{\partial}{\partial \phi_1} \spa W_2 = \frac{\partial}{\partial \phi_2} \spa W_3 = \frac{\partial}{\partial \phi_3} \spa W_4 =  \frac{\partial}{\partial t}
\end{equation}
The domains $D_1$, $D_2$ and $D_3$ corresponds to the hyperplanes of fixed points for the rotation angles $\phi_1$, $\phi_2$ and $\phi_3$, respectively, while $D_4$ corresponds to the cosmological horizon. The three domains constitutes a division of the two-sphere $S^2 = D_1 \cup D_2 \cup D_3 \cup D_4$. We have illustrated the domain structure of dS$_7$ in Figure \ref{fig:ds7d}. Using Eqs.~\eqref{7drz}, \eqref{7dbeqs} and \eqref{7dlambda} it is straightforward to compute the domain areas
\begin{equation}
| D_1 | = | D_2 | = | D_3 | = | D_4 | = \frac{1}{8} L^4
\end{equation}
as measured by \eqref{domarea}.

Turning to asymptotically dS$_7$ space-times we consider the following class of static spherically symmetric black hole space-times given by the metric \eqref{7dmet} with
\begin{equation}
\label{staticBHds7}
\begin{array}{c} \ds
 \exists \rho_0, \rho_s, \rho_c:  \rho_c> \rho_s > \rho_0  > 0 \ , \ f(\rho_0)=f(\rho_c) =0 \ , \  ( \rho^6 f)' |_{\rho=\rho_s} = 0 \ ,   \\[3mm]
 (\rho^6 f)' > 0 \ \mbox{for} \ \rho_0 \leq \rho < \rho_s \ , \  (\rho^6 f)' < 0 \ \mbox{for} \ \rho_s < \rho \leq \rho_c \ ,
 \\[3mm] \ds  f (\rho) \simeq -\frac{\rho^2}{L^2} \ \ \mbox{for}\ \  \rho \gg \rho_c  
\end{array}
\end{equation}
The choice of commuting Killing vector fields is \eqref{7dkill}. For this general class of space-times we find five domains, $D_1$ corresponding to $\theta=0$, $D_2$ to $\psi=0$, $D_3$ to $\psi=\pi/2$, $D_4$ to $\rho=\rho_0$ and $D_4$ to $\rho= \rho_c$, with the domain directions
\begin{equation}
W_1 = \frac{\partial}{\partial \phi_1} \spa W_2 = \frac{\partial}{\partial \phi_2} \spa W_3 = \frac{\partial}{\partial \phi_3} \spa W_4 =  W_5 = \frac{\partial}{\partial t}
\end{equation}
We have illustrated the domain structure of this general class of asymptotically dS$_7$, static and spherically symmetric black hole space-times in Figure \ref{fig:ds7d}.
The domains $D_1$, $D_2$ and $D_3$ correspond to hyperplanes of fixed points for the rotation angles $\phi_1$, $\phi_2$ and $\phi_3$, respectively, while $D_4$ and $D_5$ correspond to the black hole event horizon and the cosmological horizon, respectively. Using Eqs.~\eqref{7drz}, \eqref{7dbeqs} and \eqref{7dlambda} we compute the domain areas
\begin{equation}
| D_1 | = | D_2 | = | D_3 | = \frac{1}{8} ( \rho_c^4 - \rho_0^4 ) \spa | D_4 | = \frac{\rho_0^5}{16} f' (\rho_0 ) \spa | D_5 | = -\frac{\rho_c^5}{16} f' (\rho_c ) 
\end{equation}
as measured by \eqref{domarea}. Considering the specific case of a Reissner-Nordstr\" om-dS$_7$ black hole, which is a static electrically charged spherically symmetric black hole that is a solution to 7D Einstein-Maxwell gravity with a cosmological constant $\Lambda = 15/ L^2$, we have
\begin{equation}
\label{7df}
f(\rho) = 1 - \frac{\rho^2}{L^2} - \frac{2\mu}{\rho^4} + \frac{q^2}{\rho^8}
\end{equation}
and we find the domain areas
\begin{equation}
| D_1 | = | D_2 | = | D_3 | = \frac{1}{8} ( \rho_c^4 - \rho_0^4 ) \spa | D_4 | = \frac{\mu}{2} - \frac{\rho_0^6}{8L^2} - \frac{q^2}{2\rho_0^4}  \spa | D_5 | = \frac{\rho_c^6}{8 L^2} - \frac{\mu}{2} + \frac{q^2}{2 \rho_c^4}
\end{equation}

Our analysis above for spherically symmetric and static black hole space-times can be straightforwardly extended to the analysis of the domain structure of exact solutions for six- and seven-dimensional stationary black hole space-times, such as the Kerr-dS metrics of \cite{Gibbons:2004uw}. Topologically, one gets the same domain structure as found above for static and spherically symmetric black holes.

\section{Discussion and outlook}
\label{sec:concl}

In this paper we have generalized the domain structure for black hole space-times to include asymptotically dS and AdS space-times. As a step towards a full characterization of dS and AdS black holes we have found new topological and geometrical invariants of the black hole space-times. These new invariants are found by generalizing an important group of invariants known as the rod structure \cite{Harmark:2004rm,Hollands:2007aj,Hollands:2008fm} and domain structure \cite{Harmark:2009dh} found for asymptotically flat black hole space-times in four and higher dimensions. While we are not able to prove a uniqueness theorem, it is clear from our work that the new invariants are necessary in order to fully characterize dS and AdS black holes.

Our domain structure invariants have been shown in Section \ref{sec:generaldom} to be independent on which $(r,z)$ coordinate system we choose for the Canonical form of the metric \eqref{genmetric}. This means that the topological invariants in the form of the division of the domain structure into domains $D= D_1 \cup \cdots \cup D_N$, as well as the geometrical invariants which are the volumes of the domains $|D_i|$, do not depend on choices of coordinate systems. They are true invariants of the space-time. In particular, as we showed in Section \ref{sec:generaldom}, the geometrical invariants only depend on how we define $V_{(0)}$ relative to a background space-time in the asymptotic region of the space-time. This suggests that there exists a coordinate free description of the domain structure invariants for which we do not need to invoke the Canonical form of the metric \eqref{genmetric} \cite{komardomains}.

As found in Section \ref{sec:asymptads} the domain structure for asymptotically AdS black holes strongly resembles that of asymptotically flat black holes. Instead, the domain structure of asymptotically dS black holes has a more interesting structure, as seen in Sections \ref{sec:asymptds} and \ref{sec:higherdim}. Indeed, it seems the topology of the domain structure space for asymptotically dS space-times is like adding a point at infinity to the domain structure space for asymptotically flat space-times. Thus, in four and five dimensions the domain structure is $S^1$, versus $\R$ for asymptotically flat space-times, and in six and seven dimensions it is $S^2$, versus $\R^2$ for asymptotically flat space-times, here assuming stationarity and the maximal number of commuting rotational Killing vector fields. It would be interesting to examine whether this holds in higher than seven dimensions as well.   

As consequence of our domain structure analysis for stationary and asymptotically flat, dS and AdS black hole space-times with the maximal number of commuting rotational Killing vector fields, we can infer that assuming the domain structure space is connected (meaning that it is either $\R$ or $S^1$) the horizon topologies are restricted to be $S^2$ for four-dimensional black holes and $S^3$, $S^2 \times S^1$ or Lens-space topology for five-dimensional black holes.

An important application of the domain structure for asymptotically flat space-times is to the numerical study of black holes \cite{Adam:2011dn}. This is because for finding the black hole space-times numerically one needs to know what boundary conditions to put. These boundary conditions correspond to the domain structure. Thus, by extending the domain structure to asymptotically dS and AdS space-times one can use this for the numerical study of these classes of black hole space-times.

In \cite{Harmark:2009dh} various possibilities of event horizons of asymptotically flat higher-dimensional black holes based on the Blackfold method \cite{Emparan:2007wm} were examined. Similarly, one can repeat this study now for asymptotically AdS and dS black holes using the results of applying the Blackfold approach to these space-times \cite{Caldarelli:2008pz,Armas:2010hz}. Furthermore, one can easily extend this using the methods to put local and global charges on the blackfold \cite{Grignani:2010xm,Caldarelli:2010xz,Emparan:2011hg} for asymptotically flat, dS and AdS space-times. In the latter case it could be interesting to extend the invariants used in addition to the rod-structure in the uniqueness theorems for five-dimensional black holes in Einstein-Maxwell-Chern-Simons gravity \cite{Tomizawa:2009ua,Tomizawa:2009tb,Armas:2009dd}.

Finally, we remark that in \cite{Figueras:2009ci,Amsel:2009et} uniqueness theorems were proven for extremal black holes in vacuum Einstein equations, which required not only the knowledge of the conserved charges and rod-structure but also a full characterization of the near horizon geometry. This extra structure is needed to describe the black hole due to the fact that the length of the rod corresponding to the horizon KVF shrinks to zero. It would be interesting to understand how this extra structure can be generalized to asymptotically dS or AdS space-times, as well as to higher-dimensional space-times.

\section*{Acknowledgments}

TH thanks the Niels Bohr Institute for warm hospitality.
JA thanks FCT Portugal grant SFRH/BD/45893/2008 for support. JA thanks 
all the artists, musicians and audience that have set their feet at the Huset Ved Kongens Have where most of his work for this project was done and to all the scientists and cocktail geeks for their inspiring talks at the Science \& Cocktails cycle of public lectures.

\small

\providecommand{\href}[2]{#2}\begingroup\raggedright\endgroup



\end{document}